\documentclass[aps,prd,floatfix,superscriptaddress,amsmath,amssymb,twocolumn,nofootinbib]{revtex4-1}
\usepackage[dvipsnames]{xcolor}
\usepackage{graphicx}
\usepackage[colorlinks=true,citecolor=blue,linkcolor=blue,breaklinks=true]{hyperref}
\usepackage{natbib}
\usepackage{caption}
\usepackage{subcaption}
\usepackage{gensymb}

\begin{document}

\count\footins = 1000
\title{Signatures of afterglows from light dark matter boosted by supernova neutrinos in current and future large underground detectors}

\author{Yen-Hsun Lin}
\email{yenhsun@phys.ncku.edu.tw}
\affiliation{Physics Division, National Center for Theoretical Sciences, Taipei 106, Taiwan}
\affiliation{Institute of Physics, Academia Sinica, Taipei 115, Taiwan}

\author{Tsung-Han Tsai}
\email{john.9.12.5106@gmail.com}
\affiliation{Department of Physics, National Tsing Hua University, Hsinchu 300, Taiwan}


\author{Guey-Lin Lin}
\email{glin@nycu.edu.tw}
\affiliation{Institute of Physics, National Yang Ming Chiao Tung University, Hsinchu 300, Taiwan}
\affiliation{Institute of Physics, Academia Sinica, Taipei 115, Taiwan}

\author{Henry Tsz-King Wong}
\email{htwong@phys.sinica.edu.tw}
\affiliation{Institute of Physics, Academia Sinica, Taipei 115, Taiwan}

\author{Meng-Ru Wu}
\email{mwu@gate.sinica.edu.tw}
\affiliation{Institute of Physics, Academia Sinica, Taipei 115, Taiwan}
\affiliation{Institute of Astronomy and Astrophysics, Academia Sinica, Taipei 106, Taiwan}
\affiliation{Physics Division, National Center for Theoretical Sciences, Taipei 106, Taiwan}

\begin{abstract}

Supernova neutrino boosted dark matter (SN$\nu$ BDM) and its afterglow effect have been shown to be a promising signature for beyond Standard Model (bSM) physics. The time-evolution feature of SN$\nu$ BDM allows for 
possibly direct inference 
of DM mass $m_\chi$,  
and results in significant background suppression with improving sensitivity. 
This paper extends the earlier study \cite{Lin:2022dbl} and provides a general framework for computing the SN$\nu$ BDM fluxes for a supernova that occurs 
at any location in our galaxy.
A bSM $U(1)_{L_\mu-L_\tau}$ model with its gauge boson coupling to both DM and the second and third 
generation of leptons is considered, which allows for both DM-$\nu$ and DM-$e$ interactions. 
Detailed analysis of the temporal profile, angular distribution, and energy spectrum of the SN$\nu$~BDM are performed. 
Unique signatures in SN$\nu$ BDM allowing extraction of $m_\chi$ and detail features that contain information of the underlying interaction type are discussed.  
Expected sensitivities on the above new physics model from Super-Kamiokande, Hyper-Kamiokande, and DUNE detections of BDM events induced by the next galactic SN are derived and compared with the existing bounds.


\end{abstract}
\maketitle

\section{Introduction}

The nature of more than 80\% of the matter composition of the Universe -- the dark matter (DM) -- remains mysterious. 
While its presence can be inferred from the movements of stars in galaxies and the lensing of galaxy clusters due to gravitational influence, the property of DM as particles beyond the Standard Model 
remain to be discovered~\cite{Battaglieri:2017aum,Workman:2022}.
A plethora of theoretical models have been proposed to explain the origin and naturalness of DM \cite{Holdom:1985ag,He:1991qd,Davoudiasl:2012ag,Chang:2018rso,Foldenauer:2018zrz,Escudero:2019gzq,Croon:2020lrf,Lin:2021hen} but none has been proven correct as the smoking gun signatures from
both DM direct detection (DD) and indirect detection (ID) 
have yet to be observed~\cite{AMS:2015azc,LUX:2016ggv,Fermi-LAT:2017opo,LUX:2017ree,SuperCDMS:2018mne,DAMPE:2017fbg,XENON:2018voc,XENON:2019gfn,XENON:2019zpr,SENSEI:2019ibb}.
Although the DM-nucleon interaction cross section for the mass $m_\chi$ above GeV is tightly constrained 
by the DM DD~\cite{LUX:2017ree,XENON:2018voc} and approaches the neutrino floor \cite{Workman:2022},
bounds for sub-GeV DM diminishes quickly. 
On the other hand, considering the DM-electron interaction allows us to probe
much lighter DM mass down to $m_\chi\gtrsim\mathcal{O}(10)$ MeV~\cite{SENSEI:2019ibb,SuperCDMS:2018mne}.
The investigation for sub-GeV DM has gained much attention recently.

Besides the virialized DM component in the Milky Way, the halo DM can also be upscattered by high energy cosmic particles,
including nuclei, electrons and neutrinos in our Galaxy and beyond \cite{Bringmann:2018cvk,Ema:2018bih,Cappiello:2019qsw,Dent:2019krz,Wang:2019jtk,Zhang:2020nis,Guo:2020drq,Ge:2020yuf,Cao:2020bwd,Jho:2020sku,Cho:2020mnc,Jaeckel:2020oet,Lei:2020mii,Guo:2020oum,Xia:2020apm,Dent:2020syp,Ema:2020ulo,Flambaum:2020xxo,Jho:2021rmn,Das:2021lcr,Bell:2021xff,Chao:2021orr,Ghosh:2021vkt,Feng:2021hyz,Wang:2021nbf,Xia:2021vbz,Wang:2021jic,Super-Kamiokande:2022ncz,PandaX-II:2021kai,CDEX:2022fig,Granelli:2022ysi,Cline:2022qld,Xia:2022tid,Cappiello:2022exa,Carenza:2022som,Wang:2023wrx,COSINE-100:2023tcq}. 
The upscattering can possibly boost DM to a velocity $v_\chi$
close to the speed of light, much higher than that of the typical halo DM $v_\chi\sim\mathcal{O}(10^{-3})$.
As such, the boosted DM (BDM) can have large enough kinetic energy $T_\chi$ and result in detectable recoil signatures not only in current and upcoming DD experiments, but also the neutrino experiments, e.g., Super-Kamionkande (Super-K) \cite{Super-Kamiokande:2016yck}, Hyper-Kamionkande (Hyper-K) \cite{Hyper-Kamiokande:2018ofw}, DUNE~\cite{DUNE:2020ypp} and JUNO~\cite{JUNO:2021vlw}, through
the BDM interaction with targets, thereby allowing to probe much lighter DMs.

\begin{figure*}
     \centering
     \begin{subfigure}{0.3\textwidth}
         \centering
         \includegraphics{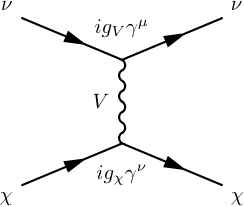}
         \caption{$\chi-\nu$ scattering}
         \label{fig:xv}
     \end{subfigure}
     \quad
     \begin{subfigure}{0.3\textwidth}
         \centering
         \includegraphics{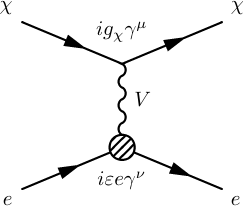}
         \caption{$\chi-e$ scattering}
         \label{fig:xe}
     \end{subfigure}
     \caption{Feynman diagrams for DM-SM interactions described by $\mathcal{L}_X$ 
     }\label{fig:xv_and_xe_digarams}
\end{figure*}

Ref.~\cite{Lin:2022dbl} recently proposed a novel BDM scenario, which considers DM upscattered by supernova neutrinos (SN$\nu$). 
For a SN explosion that happens at the center of the Milky Way or in nearby galaxies, e.g., the Large Magellanic Cloud which hosted SN1987a, the SN$\nu$ BDM flux features a unique temporal profile and can produce \emph{afterglow} events following the prompt SN$\nu$ burst. 
It was shown that the temporal features of SN$\nu$ BDM are determined mainly by $m_\chi$,  
which thus potentially allows the inference of DM mass if SN$\nu$ BDM are detected.  
The underlying concept is similar to the \emph{time of flight} (TOF) measurement for particle masses often used in laboratories, but now in astronomical scale. 
Moreover, the duration of the afterglow events are shorter for smaller $m_\chi$, which is useful in minimizing the background for deriving limits set by SN1987a event as well as for the projected sensitivities with future galactic SN explosion. 
The derived constraints and expected sensitivities on the DM--$\nu$ cross section could be complementary to other probes proposed recently~\cite{Murase:2019xqi,Dror:2019onn,Dror:2020czw,Jho:2021rmn,Das:2021lcr,Ferrer:2022kei,Mosbech:2022uud,Cline:2023tkp,Brax:2023tvn,Akita:2023yga}.


In this work, we relax two major assumptions made in Ref.~\cite{Lin:2022dbl} and perform a more thorough study for SN$\nu$~BDM from galactic SNe. 
First, we examine in detail the dependence of the temporal profile of the SN$\nu$ BDM flux on the SN location that may be far away from the galactic center (GC). 
Second, in addition to the model-agnostic case used in Ref.~\cite{Lin:2022dbl} that treats the DM--$\nu$ and DM--$e$ cross section independently, we also take specifically the 
$U(1)_{L_\mu-L_\tau}$ model with DM extension  \cite{Chang:2018rso,Croon:2020lrf,Escudero:2019gzq,Foldenauer:2018zrz}, which naturally contains  DM--$\nu$ and DM--$e$ interactions. 
As will be shown later, even for a SN located far away from the GC, the resulting SN$\nu$ BDM flux still contain distinct features that can be used as TOF measurement to infer the $m_\chi$. 
Moreover, we will show that the detailed temporal shape of the SN$\nu$ BDM also depends on the assumed underlying model. 
This means that a precise measurement of the SN$\nu$ BDM may be used to probe
different interaction types, eg.~scalar, vector, axial-vector,...etc.


We begin our paper by introducing the extended $U(1)_{L_\mu-L_\tau}$ model and derive the scattering amplitude of DM--$\nu$ and DM--$e$ and the interaction cross sections in Sec.~\ref{sec:pheno_setup}. 
In Sec.~\ref{sec:flux}, we present the generalized formalism to compute the time-dependent flux of SN$\nu$ BDM for a SN at any location in the Milky Way.
Sec.~\ref{sec:event} discusses features in derived SN$\nu$ BDM fluxes for different SN locations and compare results obtained with the extended $U(1)_{L_\mu-L_\tau}$ model to those obtained with a model-agnostic approach~\cite{Lin:2022dbl}. 
Projected sensitivities for the considered models are given in Sec.~\ref{sec:sensitivity}, before we summarize in Sec.~\ref{sec:summary}. 
To maintain the structure of the paper, we leave all derivations to the appendices.

\section{Phenomenological setup}\label{sec:pheno_setup}

There are various ways to introduce DM interaction with Standard Model (SM) particles based on an effective Lagrangian or a phenomenological model construction. 
For having both DM-$\nu$ and DM-$e$ interactions, existing phenomenological models such as the $Z$-mass mixing \cite{Davoudiasl:2012ag,Lin:2021hen} and $U(1)_{L_\mu-L_\tau}$ \cite{Chang:2018rso,Croon:2020lrf,Escudero:2019gzq,Foldenauer:2018zrz} (either with a non-zero kinetic mixing \cite{Holdom:1985ag} term that couples DM and $e$ directly or not) models provide the portals. 
In $U(1)_{L_\mu-L_\tau}$ model,   
DM couples to neutrinos through a 
gauge boson $V$, which connects the 2nd- and 3rd- generation leptons and neutrinos to the dark sector. 
The relevant part of the Lagrangian can be written as 
\begin{alignat}{1}
    \mathcal{L}_\chi  \supset & -\frac{1}{4}V_{\mu\nu}V^{\mu\nu}+\frac{\varepsilon}{2\cos\theta_W}F_{\mu\nu}V^{\mu\nu}-\frac{1}{2}m_V^2 V_\mu V^\mu \nonumber \\
    & \quad -m_\chi \bar{\chi}\chi +g_\chi V_\mu \bar{\chi}\gamma^\mu \chi \nonumber \\
    & \quad +g_V V_\mu Q_{\alpha\beta}\left(\bar{\ell}_\alpha\gamma^\mu \ell_\beta +\bar{\nu}_\alpha \gamma^\mu P_L \nu_\beta\right),\label{eq:Lagrangian}
\end{alignat}
where $V_{\mu\nu} =\partial_\mu V_\nu -\partial_\nu V_\mu$ and $F_{\mu\nu}$ are
the $U(1)_{L_\mu-L_\tau}$ and SM $U(1)_Y$ field strength tensors respectively, $\chi$ the fermionic DM field with mass $m_\chi$ and the $U(1)_{L_\mu-L_\tau}$ charge $g_{\chi}$,
$m_V$ the $V$ boson mass, and $\varepsilon$ the kinetic mixing parameter.  
The SM charged lepton and neutrino fields are specified by $\ell_\alpha$ and $\nu_\alpha$ where $\alpha=e, \mu,\tau$ and the matrix $Q_{\alpha\beta}\equiv {\rm diag}(0,1,-1)$ gives the $U(1)_{L_\mu-L_\tau}$ charges of leptons and neutrinos. 
In this model, the strengths of DM-$\nu$ and DM-$e$ interaction vertices in the tree level are determined by $g_V$ and $\varepsilon e$, respectively. 
The corresponding Feynman diagrams are displayed in Fig.~\ref{fig:xv_and_xe_digarams}.
Obviously, this model also leads to DM self-interaction mediated by $V$ with coupling strength $g_\chi$. 

\begin{figure}
\begin{centering}
\includegraphics[width=0.8\columnwidth]{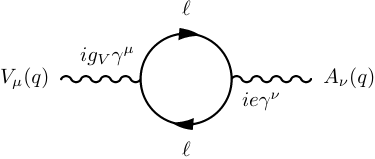}
\end{centering}
\caption{\label{fig:mutau_loop}
SM particle carries EM charge $e$ can couple to $V_\mu$ through the loop effect with $\ell=\mu,\tau$.
}
\end{figure}

Interestingly, non-vanishing DM-$e$ interaction can also be generated through a loop-induced effective mixing even though $\varepsilon=0$. 
As seen in Fig.~\ref{fig:mutau_loop}, any SM particle carrying EM charge can couple to $V_\mu$ via a $\mu/\tau$-loop. 
For $\varepsilon\neq 0$, this effect is generally suppressed and therefore negligible. 
However, it becomes the leading order contribution to DM-$e$ interaction for $\varepsilon=0$ and its strength is given 
by the induced kinetic mixing parameter~\cite{Chang:2018rso,Escudero:2019gzq}
\begin{equation}
    \varepsilon^\prime =  -\frac{g_V}{2\pi^2}\int_0^1 dx\,x(1-x)\ln\left[\frac{m_\tau^2-x(1-x)q^2}{m_\mu^2-x(1-x)q^2}\right],\label{eq:induced_eps}
\end{equation}
where $q^{\mu}$ is the $4$-momentum transfer. 
This effective mixing can be approximated as   $\varepsilon^\prime/g_V\approx -1/70$ for cases with $m^2_\mu \gg -q^2$.

Given this Lagrangian, we can compute the corresponding $\chi-e$ and $\chi-\nu$ cross sections through the following scattering amplitudes
associated with the Feynman diagrams shown in Fig.~\ref{fig:xv_and_xe_digarams},
\begin{align}
|\mathcal{M}|^{2} & =2\left(\frac{\mathcal{Q}}{t_{M}-m_{V}^{2}}\right)^{2}[s_{M}^{2}+u_{M}^{2}\nonumber \\
 & \quad +4t_{M}(m_{1}^{2}+m_{2}^{2}) -2(m_{1}^{2}+m_{2}^{2})^{2}], \label{eq:amplitude}
\end{align}
where $m_{1,2}$ are the masses of two particles in the initial state and $(s_M,t_M,u_M)$ the Mandelstam variables.
For processes depicted in Figs.~\ref{fig:xv} and \ref{fig:xe}, $\mathcal{Q}=g_V g_\chi$ and $g_\chi \varepsilon e$ respectively.
We give all the derivation details, the resulting cross sections, and relevant discussions in Appendix~\ref{appx:model_kinematics}. 

As can be seen from Appendix~\ref{appx:model_kinematics}, the $\chi-e$ and $\chi-\nu$ cross sections, $\sigma_{\chi\nu}$ and $\sigma_{\chi e}$, are energy dependent, which is originated from the assumed interaction type 
in $\mathcal{L}_\chi$, in contrast to the model-agnostic case assumed in \cite{Lin:2022dbl} where these cross sections are assumed to be energy-independent. 
For all results computed with the extended $U(1)_{L_\mu-L_\tau}$ model, specific model parameters $g_\chi$, $g_V$, $\varepsilon$ and $m_V$ will be given. 
On the other hand, those results derived with the model-agnostic approach would only depend on the values of cross sections $\sigma_{\chi\nu}$ and $\sigma_{\chi e}$. 

We note that Eq.~\eqref{eq:Lagrangian} also gives rise to neutrino non-standard self-interaction ($\nu$NSI), via the exchange of gauge boson $V$, which may affect the SN$\nu$ emission at the source~\cite{Chang:2022aas,Fiorillo:2023ytr} or alter the SN$\nu$ spectra during their propagation by interacting with the cosmic neutrino background~\cite{Shalgar:2019rqe}. 
For the latter, it requires a much larger $g_V$ than what we explore in this paper to significantly distort the SN$\nu$ spectra during their propagation and can be safely ignored.  
For the former, it may lead to a different decoupling behavior of SN$\nu$ but the effect on SN$\nu$ emission remain debated~\cite{Chang:2022aas,Fiorillo:2023ytr}.
Thus, we ignore this effect in the rest of this paper.

\section{Framework for computing the SN$\nu$ BDM flux}\label{sec:flux}

\subsection{The geometry}

\begin{figure}
\begin{centering}
\includegraphics[width=0.7\columnwidth]{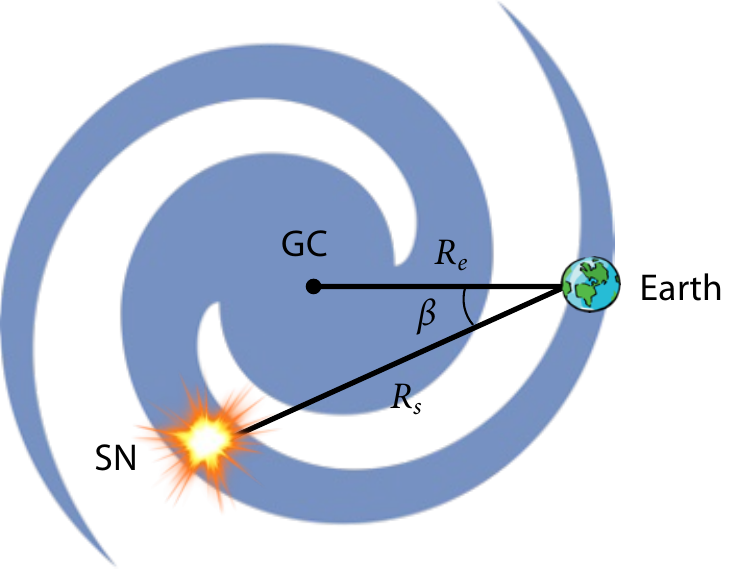}
\end{centering}
\caption{\label{fig:geometry_topView}
The top-view of GC-SN-Earth system where the three locations 
automatically form a plane.}
\end{figure}
In this section, we follow Ref.~\cite{Lin:2022dbl} and consider DM boosted by neutrinos emitted from a single SN explosion. As a generalization of~\cite{Lin:2022dbl}, the SN can be located anywhere in the galaxy, instead of being restricted to the position of the GC. 
Given that the three locations, GC, SN and Earth, lie on the same plane shown in Fig.~\ref{fig:geometry_topView}, we define $R_e$ the distance between Earth and GC, $R_s$ the distance from Earth to SN, and $\beta$ the angle between the directions of the GC and the SN viewed from Earth, which characterizes how far the SN is away from the direction of GC. 

\begin{figure}
\begin{centering}
\includegraphics[width=0.8\columnwidth]{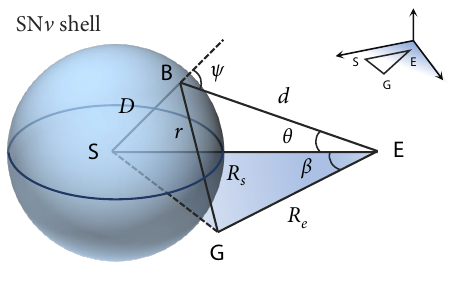}
\end{centering}
\caption{\label{fig:geometry_scheme}
The 3D schematic diagram of SN$\nu$ BDM.
The SN$\nu$ shell indicates the outward propagation of SN$\nu$ from the SN explosion
at $\mathsf{S}$.
}
\end{figure}

To depict DM boosted by neutrinos emitted from the SN, we rely on the three dimensional (3D) geometry shown in Fig.~\ref{fig:geometry_scheme}. 
In this figure, we denote GC, SN and Earth as $\mathsf{G}$, $\mathsf{S}$ and  $\mathsf{E}$. 
We approximate that SN neutrinos are emitted from the SN location $\mathsf{S}$ within a duration $\tau_s=10$~s and form a spherical thin shell with a fixed width $c\tau_s\ll \{ R_s, R_e\}$, which expands radially with an increasing radius $D$. 
For DM being scattered off at the location $\mathsf{B}$ and travel a distance $d$ to the Earth $\mathsf{E}$, it requires a scattering angle $\psi$ relative to the normal direction of the SN$\nu$ shell, with $\theta$ labeling the angle between the line connecting $\mathsf{S}$ and $\mathsf{E}$, and that connecting $\mathsf{B}$ and $\mathsf{E}$. 
The local DM density at $\mathsf{B}$ is determined by the DM halo profile $n_\chi(r)$, taken to be spherically symmetric with respect to the GC $\mathsf{G}$, where $r$ denotes the distance from $\mathsf{B}$ to $\mathsf{G}$.
Notice that we have neglected the motion of the Earth relative to GC and taken $R_e=8.5$~kpc as a constant.
This is well justified since the rotation velocity of the Solar system relative to GC is $v_\odot\sim 255$~km~s$^{-1}$, which is much smaller than the velocities of SN$\nu$ and the boosted DM of our interest, both of which are close to the speed of light\footnote{For example, within a typical time scale defined by the traveling time of SN$\nu$ arriving the Earth, from GC, $t_c\equiv R_e/c$, the Earth only moves a distance $l \approx v_\odot t_c \approx 6.2\times 10^{-3}$~kpc, which is obviously much smaller than $R_e$.}.
Also noted is that in the following we take $R_s$ and $\beta$ as known values for demonstrating the dependence of the results on the SN location\footnote{  
In fact, we expect that when the next Galactic SN occurs, the detection of its electromagnetic emission, neutrinos, as well as gravitational waves will help pin down these parameters to a good precision.}.

\begin{figure*}
\begin{centering}
\includegraphics[width=0.8\textwidth]{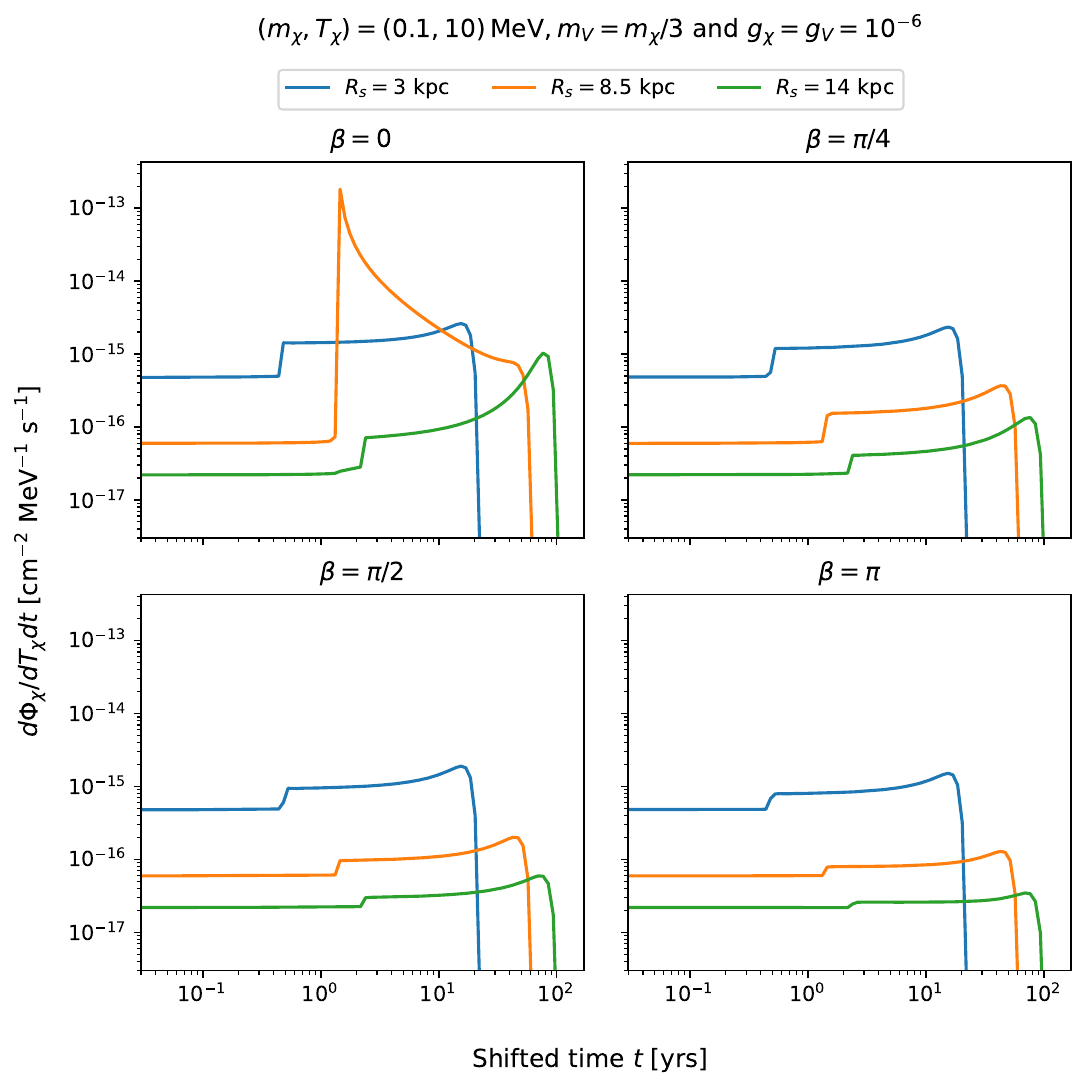}
\end{centering}
\caption{The SN$\nu$ BDM flux on Earth with different $\beta$ (each subplots) and $R_s$ (colors). Dark sector parameters
for computing these plots are labeled on top of the figure. The temporal axis (horizontal) is adjusted to shifted coordinate $t$ and displayed in yrs with $t=0$ indicates the arrival time of SN$\nu$ on Earth.}
\label{fig:flux_1MeV_md}
\end{figure*}

\subsection{The BDM emissivity and the flux on the Earth}

To calculate the SN$\nu$ BDM flux at Earth, we first evaluate the BDM emissivity
$j_\chi$ at $\mathsf{B}$ depicted in Fig.~\ref{fig:geometry_scheme}. 
The emissivity can be written down as (see also~\cite{Lin:2022dbl})
\begin{equation}
j_{\chi}(r,D,T_{\chi},\psi)=cn_{\chi}(r)
\left(\frac{1}{2\pi}\frac{d\sigma_{\chi\nu}}{d\cos\psi}\right)
\left(\frac{dn_{\nu}}{dE_{\nu}}\right) \left(\frac{dE_{\nu}}{dT_{\chi}}\frac{v_\chi}{c}\right), 
\label{eq:jx}
\end{equation}
where we take the NFW profile~\cite{Navarro:1995iw, Navarro:1996gj, Bertone:2004pz} for $n_\chi(r)$. 
The factor $dE_\nu/dT_\chi$ and the differential DM-$\nu$ cross section $d\sigma_{\chi\nu}/d\cos\psi$ are given by Eqs.~\eqref{eq:dEv/dTx} and \eqref{eq:diff_sigxv_psi} respectively for the $U(1)_{L_\nu-L_\tau}$ model.  
For the model-agnostic case, we take $d\sigma_{\chi\nu}/d\cos\psi=\sigma_{\chi\nu}\times f_\chi(\psi)$ where $f_\chi(\psi)= \gamma^2\sec^{3}\psi/(\pi(1+\gamma^{2}\tan^{2}\psi)^{2})$ with $\gamma=(E_\nu+m_\chi)/\sqrt{m_\chi(2E_\nu+m_\chi)}$ \cite{Lin:2022dbl}.

For the SN$\nu$ number density, we take the same expression as in Ref.~\cite{Lin:2022dbl}, 
\begin{equation}
\frac{dn_{\nu}}{dE_{\nu}} =\sum_{i}\frac{L_{\nu_{i}}}{4\pi D^{2}\langle E_{\nu_{i}}\rangle}E_{\nu}^{2}f_{\nu_{i}}(E_{\nu}), \label{eq:dnv/dEv}
\end{equation}
where $L_{\nu_{i}}=L_{\nu,{\rm tot}}/6$ 
is the luminosity 
of each flavor ($\nu_e$,~$\nu_\mu$,~$\nu_\tau$ and their antineutrinos).  
The average energy $\langle E_{\nu_e}\rangle$,~$\langle E_{\bar\nu_e}\rangle$, 
and~$\langle E_{\nu_x}\rangle$ ($\nu_x\in \{ \nu_\mu, \nu_\tau, \bar\nu_\mu, \bar\nu_\tau\}$) are taken to be 11,~16,~25~MeV, respectively~\cite{Duan:2006an}. 
We assume a Fermi-Dirac distribution $f_{\nu_{i}}$ with a pinch parameter $\eta_{\nu_i}\equiv\mu_{\nu_i}/T_{\nu_i}=3$, such that $T_{\nu_i} \approx \langle E_{\nu_i}\rangle/3.99$. 

Given the emissivity, one can then integrate it over the solid angle spanned by $\theta$ and $\phi$ (viewed from $\mathsf{E}$ in spherical coordinate) to obtain the SN$\nu$ BDM flux on Earth at time $t^\prime$ after the SN explosion as 
\begin{widetext}
\begin{equation}
\frac{d\Phi_{\chi}(T_\chi, t^\prime)}{dT_{\chi}dt} =
\left.\tau_s\int_0^{2\pi} d\phi\int_{0}^{\pi/2}\sin\theta d\theta~ \mathcal{J} j_{\chi}(r(\phi),D,T_{\chi},\psi)\right|_{t^{\prime}=\frac{D}{c}+\frac{d}{v_{\chi}}},\label{eq:BDM_flux}
\end{equation}
\end{widetext}
where the Jacobian reads
\begin{equation}
    \mathcal{J}=\left(\frac{d-R_s\cos\theta}{cD}+\frac{1}{v_{\chi}}\right)^{-1},\label{eq:J}
\end{equation}
Note that different from Ref.~\cite{Lin:2022dbl}, the integration over the azimuthal angle $\phi$ needs to be carried out explicitly. 
This is because when SN is not located at the GC,
the DM number density is not spherically symmetric with respect to $\mathsf{S}$ (see Fig.~\ref{fig:geometry_scheme}).
Also noted is that the relation
\begin{equation}
    t^\prime = \frac{D}{c}+\frac{d}{v_\chi}\label{eq:t_prime}
\end{equation}
represents the total propagation time of SN$\nu$ reaching $\mathsf{B}$ plus the time that the BDM takes from $\mathsf{B}$ to $\mathsf{E}$ (see Fig.~\ref{fig:geometry_scheme}). 
That is, to evaluate the BDM flux on the Earth at $t^\prime$,
the lengths $D$ and $d$ must satisfies Eq.~\eqref{eq:t_prime}.
The BDM velocity is
\begin{equation}
    v_\chi = \frac{\sqrt{T_\chi (2m_\chi+T_\chi)}}{m_\chi+T_\chi}c.\label{eq:vx}
\end{equation} 
We can also shift $t^\prime$ by subtracting out the constant factor $t_\nu\equiv R_s/c$, which is the propagation time
of SN$\nu$ from $\mathsf{S}$ to $\mathsf{E}$.
This shifted time coordinate $t=t^\prime-t_\nu$ is the delayed arrival time for BDM relative to
SN$\nu$ burst observed on Earth.
We give the detailed expressions that one can use to practically evaluate Eq.~\eqref{eq:BDM_flux} through the constraint Eq.~\eqref{eq:t_prime} in Appendix~\ref{appx:geometry}


\section{SN$\nu$ BDM flux on Earth}\label{sec:event}

\begin{figure*}
\begin{centering}
\includegraphics[width=0.8\textwidth]{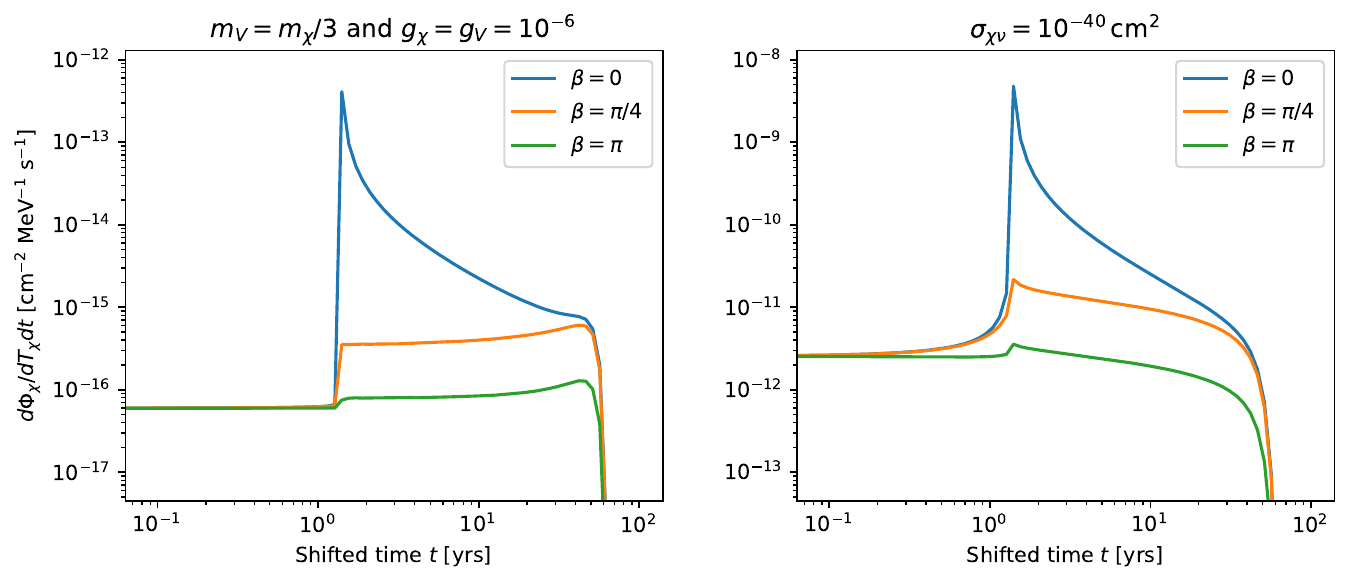}
\end{centering}
\caption{Comparison of the SN$\nu$ BDM flux obtained with the $U(1)_{L_\mu-L_\tau}$~(left) and with the model-agnostic cross sections (right) for $R_s=8.5$~kpc and different $\beta$. The DM kinetic energy  $T_\chi=10\,$ MeV and mass $m_\chi =0.1\,$MeV.
}
\label{fig:flux_v_beta}
\end{figure*}

In this section, we show the SN$\nu$ BDM fluxes computed based on Eq.~\eqref{eq:BDM_flux} on Earth. 
In Sec.~\ref{sec:res_SN_loc}, we examine the dependence of the flux on the SN location. 
In Sec.~\ref{sec:res_model}, we compare the temporal profile of the fluxes derived with $U(1)_{L_\mu-L_\tau}$ and model-agnostic cross sections. 
We then discuss the $m_\chi$ dependent general feature contained in the flux that can in principle be used to infer the value of $m_\chi$ in Sec.~\ref{sec:res_mchi}. 
In Secs.~\ref{sec:res_ang} and \ref{sec:res_energy}, we further investigate the detailed angular distribution and the energy spectrum of the BDM.


\subsection{Dependence on SN location}
\label{sec:res_SN_loc}
For the $U(1)_{L_\mu-L_\tau}$ model, we show in Fig.~\ref{fig:flux_1MeV_md} the SN$\nu$ BDM flux obtained with $(m_\chi,T_\chi)=(0.1,10)\,{\rm MeV}$, $m_V=m_\chi/3$, and $g_V=g_\chi=10^{-6}$ as a function of the shifted time $t$ (relative to the arrival time of the SN$\nu$) for cases where the SN occur at different locations with different values of SN distance $R_s$ and off-center angle $\beta$ (see Fig.~\ref{fig:geometry_topView}). 
The top-left, top-right, bottom-left, and bottom-right panels correspond to results with $\beta=0$, $\pi/4$, $\pi/2$, and $\pi$, respectively. 
Each panel contains fluxes calculated with three different values of $R_s=3$~kpc (blue), $8.5$~kpc (orange), and $14$~kpc (green). 
Clearly, the SN$\nu$ BDM flux contain interesting time-dependent features, which can be summarized as follows.

First, for all cases, the fluxes initially stay nearly constant at smaller $t$, and experience a step-wise or sharp increase at a specific moment, dubbed as $t_p$, which is independent of the angle $\beta$. 
For $R_s=3$, 8.5 and 14~kpc, $t_p\approx 0.41$, 1.4, and 2.3~yrs, respectively.
These values of $t_p$ correspond precisely to the traveling time of BDM from the SN location to Earth relative to that of neutrino, i.e., 
\begin{equation}\label{eq:tp}
t_p=\frac{R_s}{v_\chi}-t_\nu.
\end{equation}
The sharp increase of the BDM flux at $t_p$, even for SN that is away from the GC ($\beta>0$), can be primarily attributed to the increase of $dn_\nu/dE_\nu$ as $D\to0$ [see Eq~\eqref{eq:dnv/dEv}]\footnote{We checked numerically that this feature no longer exists if we artificially remove the $D$-dependence in $dn_\nu/dE_\nu$.}.
Note that for the case where the SN is located at the GC ($R_s$=8.5~kpc and $\beta=0$), the cuspy profile of DM near GC results in a much greater rise of the BDM flux as discussed in Ref.~\cite{Lin:2022dbl} than other cases where the SN are off-GC.
In order to prevent divergence of both $dn_\nu(D)/dE_\nu$ and $n_\chi(r)$ as $D$ and $r$ approach zero, we impose a cutoff of $10^{-5}$ kpc on both parameters. Any contributions below this distance are ignored. We note that even if the cutoff is smaller than $10^{-5}$ kpc, the numerical results will remain largely unaffected.

Second, for all cases, the BDM fluxes for a given $T_\chi$ vanish at times $t_{\rm van}\approx 24$, 60, and 101~yrs for $R_s=3$, 8.5, and 14~kpc. 
Similar to $t_p$, the value of $t_{\rm van}$ does not depend on $\beta$ either. 
The reason is related to the geometry shown in Fig.~\ref{fig:geometry_scheme} and that there exists a maximal scattering angle $\psi_{\rm max}$ for BDM due to the kinematic constraint (see Appendix~\ref{appx:model_kinematics}).
Since $t_{\rm van}$ represents the maximally possible traveling time for the BDM, it happens when $\psi=\psi_{\rm max}$ and when
$t^\prime = D(\theta)/c+d(\theta)/v_\chi$
takes its maximal value.
By taking $(dt^\prime/d\theta)|_{\theta=\theta^*}=0$, it is straightforward to show that this happens when $\theta=\theta^*$, which satisfies
\begin{equation}
\frac{\cos(\psi_{\rm max}-\theta^*)}{\cos\theta^*} = \frac{v_\chi}{c}  \label{eq:theta*}
\end{equation}
such that 
\begin{equation}\label{eq:tvan}
    t_{\rm van} = \frac{D(\theta^*)}{c}+\frac{d(\theta^*)}{v_\chi}-t_\nu.
\end{equation}
Clearly, both $t_p$ and $t_{\rm van}$ scales linearly with $R_s$, which can also be seen from Fig.~\ref{fig:flux_1MeV_md}. 
In fact, when taking the approximation of $m_\chi / T_\chi\ll 1$, one can further show that 
\begin{equation}\label{eq:tp_tvan_app}
 t_p \simeq  \frac{m_\chi^2}{2T_\chi^2} \frac{R_s}{c},~~ t_{\rm van} \simeq  \frac{m_\chi}{4T_\chi} \frac{R_s}{c}.    
\end{equation}
Thus, for a given $m_\chi$ ($T_\chi$), $t_p$ and $t_{\rm van}$ are larger (smaller) for a smaller $T_\chi$ ($m_\chi)$.   
The presence of $t_p$ and $t_{\rm van}$ in the BDM flux plays an important role for potentially inferring $m_\chi$ from future detection as well as for reducing the background, as will be further discussed later.

For cases with non-zero $\beta$, the BDM flux are generally larger for smaller $R_s$, which is mainly related to the higher SN$\nu$ number density that enters the BDM emissivity [cf.~Eqs.~\eqref{eq:jx} and \eqref{eq:dnv/dEv}].  
The local DM density along the path from SN to Earth only has a subdominant impact, with smaller $\beta$ giving rise to a bit larger BDM flux after $t_p$.  
In between $t_p$ and $t_{\rm van}$, the BDM flux increase monotonically, which is related to the assumed underlying particle physics model and will be discussed in the next subsection.

The cases with $\beta=0$ and with $R_s\gtrsim R_e=8.5$~kpc (upper left panel) are special since the SN$\nu$ can upscatter DM around GC where the NFW profile peaks. 
When $R_s=R_e$, the BDM flux has a large peak at $t_p$ and decreases afterwards. 
For $R_s>R_e$, the increase of the BDM flux in between $t_p$ and $t_{\rm van}$ is more pronounced than the $R_s< R_e$ case, 
due to both the larger SN$\nu$ fluxes as well as the higher DM density approaching the GC.

\begin{figure*}
\begin{centering}
\includegraphics[width=0.8\textwidth]{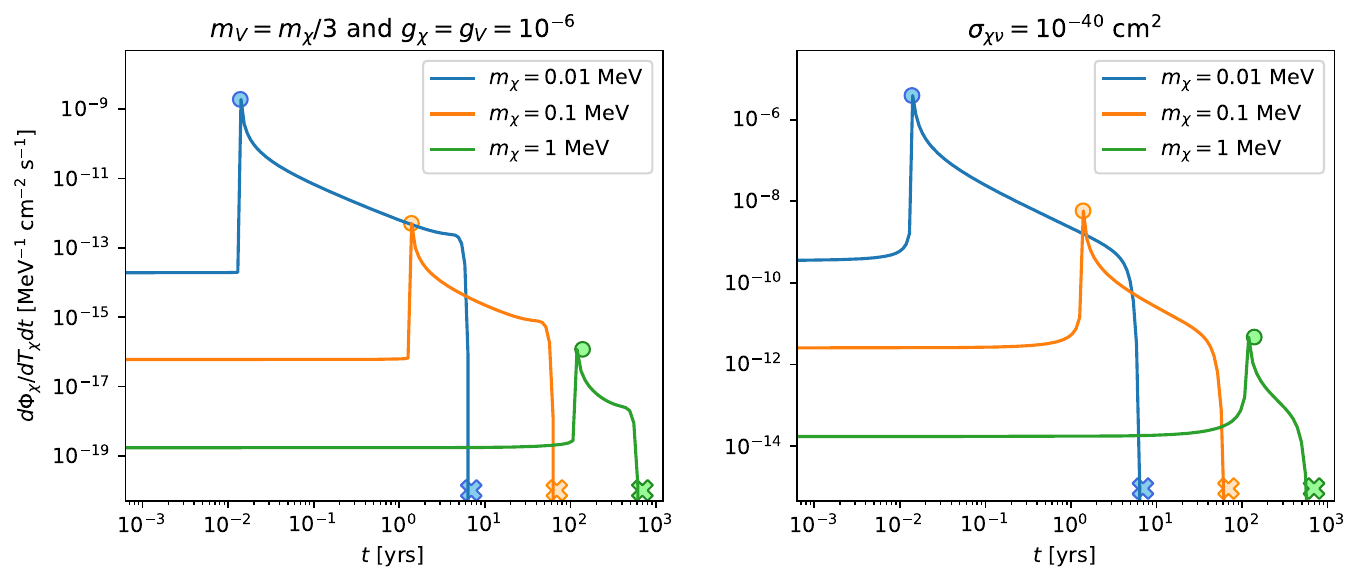}
\end{centering}
\caption{SN$\nu$ BDM flux with different $m_\chi$ for $U(1)_{L_\mu-L_\tau}$ (left) and model-agnostic cross section (right) with $R_s=8.5$~kpc and $\beta=0$. The filled circles and crosses are $t_p$ and $t_{\rm van}$, respectively, obtained from Eq.~\eqref{eq:tp_tvan_app}. These approximated values apply to both plots, which demonstrate BDM time-dependent features do not depend on the chosen model.
}
\label{fig:flux_v_mx}
\end{figure*}

\subsection{Dependence on particle physics model}
\label{sec:res_model}

We show in Fig.~\ref{fig:flux_v_beta} the comparison of the BDM fluxes obtained with $(m_\chi,T_\chi)=(0.1,10)$~MeV,  $R_s=8.5$~kpc and $\beta=0,~\pi/4$, and $\pi$, respectively, for $U(1)_{L_\mu-L_\tau}$ model (left panel) and the model-agnostic scenario (right panel). 
For $U(1)_{L_\mu-L_\tau}$ model, we use the same model parameters adopted earlier, while for the model-agnostic scenario, we assume a total cross section $\sigma_{\chi\nu}=10^{-40}$~cm$^2$.
The choice of this value manifests the current limit on the DM-$e$ cross section as we generally assume $\sigma_{\chi\nu}=\sigma_{\chi e}$ for model-agnostic case.
Fig.~\ref{fig:flux_v_beta} suggests that for most SN locations, the general presence of $t_p$ and $t_{\rm van}$ persists, insensitive to the choice of the model.  
In fact, we find that only when the SN takes place at $R_s\gtrsim 11$~kpc with $\beta\lesssim 0.02\pi$, the steeply-rising feature of the flux happens earlier than $t_p$ for the model-agnostic scenario. 
Detail discussions and figures are given in Appendix~\ref{appx:tp_model_agnostic}.

Interestingly, the temporal profiles of the fluxes in between $t_p$ and $t_{\rm van}$ differ for these two models, especially for case with non-zero $\beta$.  
For both $\beta=\pi/4$ and $\pi$, the BDM fluxes keep increasing after $t_p$ before reaching their maximal shortly before $t_{\rm van}$ for the $U(1)_{L_\mu-L_\tau}$ model. 
On the other hand, the fluxes decrease monotonically in the corresponding cases for the energy-independent model.  
This is understood that for any BDM arriving in between $t_p$ and $t_{\rm van}$, the latter the BDM arrives the larger the scattering angle $\psi$ is.  
Kinematically, a larger $\psi$ with fixed $T_\chi$ requires a larger incoming $E_\nu$ as shown by Eq.~\eqref{eq:Ev}.
Consequently, the differential cross section $d\sigma_{\chi\nu}/d\cos\psi$ under $U(1)_{L_\mu-L_\tau}$ model is also larger due to the energy-dependent nature of the cross section [cf.~Eqs.~\eqref{eq:amplitude} and \eqref{eq:diff_sigxv_psi} and the discussion in Appendix~\ref{appx:model_kinematics}]. 
As seen on the left panel of Fig.~\ref{fig:flux_v_beta}, such a cross section enhancement leads to a mild  increases of BDM flux with time for $t>t_p$ until the point that the required neutrino energy becomes too high so that the exponential suppression in the SN$\nu$ distribution [cf.~Eq.~\eqref{eq:dnv/dEv}] sets in. 

For the model-agnostic case, the differential cross section remains constant even for a scattering angle $\psi$ that requires a larger $E_\nu$. 
As a result, the BDM fluxes decrease shortly after $t_p$ for non-zero $\beta$, as shown on the right panel of Fig.~\ref{fig:flux_v_beta}. 
The dependency of the exact temporal profile of BDM flux on the chosen interaction model suggests the possibility of determining the energy dependency of DM-neutrino interaction cross section, provided the above temporal profile can be precisely measured.   

\subsection{Inferring $m_\chi$ from SN$\nu$ BDM measurement}
\label{sec:res_mchi}

\begin{figure*}
\begin{centering}
\includegraphics[width=0.8\textwidth]{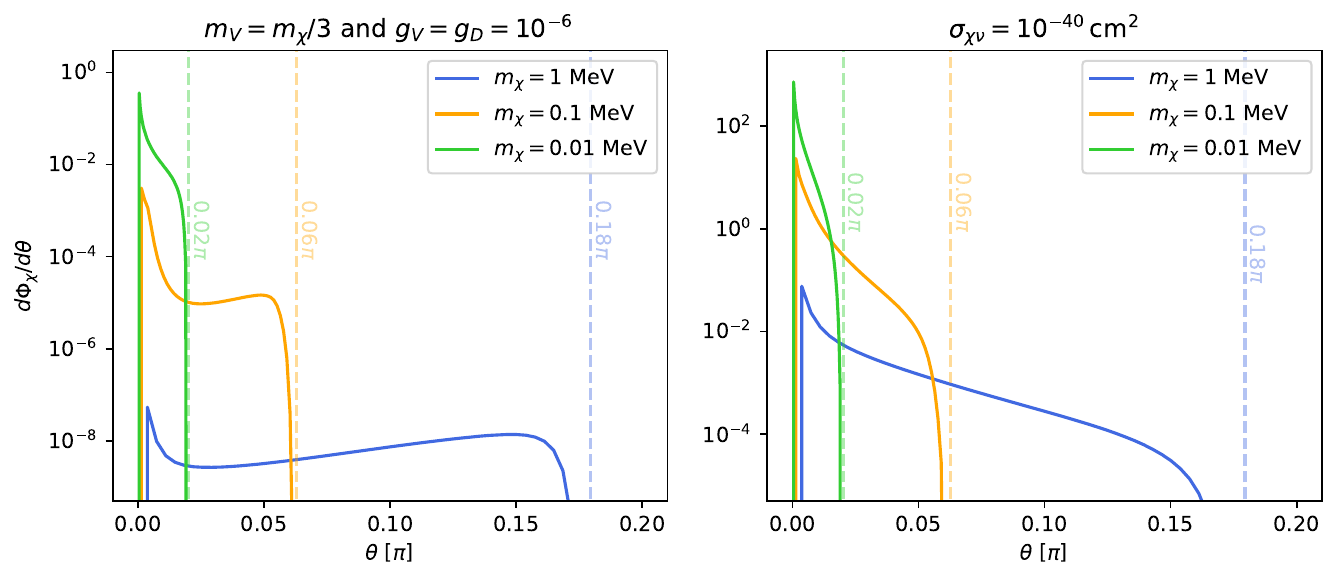}
\end{centering}
\caption{The angular distribution of the SN$\nu$BDM  versus $\theta$ [see Eq.~\eqref{eq:dPhi/dtheta}] for $U(1)_{L_\mu-L_\tau}$ (left) and model-agnostic (right) $\sigma_{\chi e}$. Different colors represent different choice of $m_\chi$.
Dashed lines are the the maximum open angle $\theta_{\rm max}\approx 0.18\pi,0.06\pi$ and $0.02\pi$ for $m_\chi=1,0.1$ and $0.01$ MeV, respectively.}
\label{fig:event_per_cos_theta}
\end{figure*}

Based on the results shown in the previous subsections, it is clear that the temporal profile of SN$\nu$ BDM flux contain two unique features -- the rapidly increase of the flux around $t_p$ for most SN locations (except $R_s\gtrsim 11$~kpc and $\beta\lesssim 0.02\pi$), and the termination of flux at $t_{\rm van}$.  
Since both $t_p$ and $t_{\rm van}$ only depend on $m_\chi$, $T_\chi$, and $R_s$ [see Eq.~\eqref{eq:tp_tvan_app}], if the number of detected SN$\nu$ BDM events is sufficient for identifying $t_p$  and $t_{\rm van}$ for any given $T_\chi$, one can directly infer $m_\chi$ provided  $R_s$ (the SN location) can be determined from other SN multimessenger signals.

To further demonstrate the dependence of the BDM flux on $m_\chi$, we show in Fig.~\ref{fig:flux_v_mx} the fluxes obtained with different $m_\chi$ values for $R_s=8.5$~kpc and $\beta=0$. 
The left panel is for the $U(1)_{L_{\mu}-L_\tau}$ and the right panel for the model-agnostic scenario. 
The values of $t_p$ and $t_{\rm van}$ computed using Eq.~\eqref{eq:tp_tvan_app} are indicated by filled circles and crosses on both panels. 
For $m_\chi=0.01$, 0.1, and 1~MeV, 
Eq.~\eqref{eq:tp_tvan_app} gives 
$t_p=0.013$, 1.38, and 138 yrs and $t_{\rm van}=6.93$, 69.3, and 693~yrs, respectively.
Clearly, the BDM fluxes between $t_p$ and $t_{\rm van}$ arrive later for larger $m_\chi$ in the same way.  
and the above values are good approximations to what obtained numerically in both scenarios.
We note here that the underlying concept of such possible inference of $m_\chi$ is the same as the TOF measurement used in laboratory to infer particle's mass, which can be clearly seen in Eqs.~\eqref{eq:tp} and \eqref{eq:tvan}, but now carried over to the astronomical scale for BDM. 
In fact, similar concepts are often used to constrain neutrino and graviton masses~\cite{Arnett1987,Kolb1987,deRham:2016nuf}.

Although the above proposal of using $t_p$ and $t_{\rm van}$ to infer $m_\chi$ is independent of the particle physics model and the cross section for most SN locations, the intensity of the BDM flux is obviously proportional to the strength of the interaction. Moreover, it is also clear from the discussion in Sec.~\ref{sec:res_model} that the detailed temporal profile of the BDM fluxes can potentially be used to infer the energy dependence of the cross section. 
Thus, the detection of the SN$\nu$ BDM events has the potential to probe various properties of the unknown dark sector, owing to the rich information encoded in the temporal profile of the BDM fluxes\footnote{In contrast, other important and complementary BDM scenarios often involve stationary sources to upscatter the DM, such that the resulting fluxes lack the time-dependent features discussed here, which make it more difficult to unfold the degeneracy between $m_\chi$ and the cross sections. }.

\subsection{BDM angular distribution}
\label{sec:res_ang}

Since the SN$\nu$ BDM generally have finite scattering angle $\psi>0$, it is also of interest to examine their angular distribution on Earth. 
For this purpose, we compute the BDM flux integrated over $T_\chi$, $t$, and the azimuthal angle $\phi$ 
\begin{widetext}
\begin{equation}
    \frac{d\Phi_\chi(\theta)}{d\theta} =  
    \int_{t_0}^{t_{\rm exp}} dt 
    \int_{T_{\chi,{\rm min}}}^{T_{\chi,{\rm max}}}dT_\chi 
    \int_0^{2\pi} d\phi\, \sin\theta
    \left.\tau_s \mathcal{J} j_{\chi}(r(\theta,\phi),D(\theta),T_{\chi},\psi)\right|_{t^{\prime}=\frac{D}{c}+\frac{d}{v_{\chi}}},\label{eq:dPhi/dtheta}
\end{equation}
\end{widetext}
with $(T_{\chi,{\rm min}},T_{\chi,{\rm max}})=(5,100)$~MeV, 
$t_0=10\,{\rm s}$, and $t_{\rm exp}={\rm min}(t_{\rm van},t_{\rm cut})$, where $t_{\rm cut}=35$~yrs.
The results are shown in Fig.~\ref{fig:event_per_cos_theta} for the $U(1)_{L_\mu-L_\tau}$ model (left panel) and for the model-agnostic case (right panel) with model parameters given explicitly on each panel. 
For both cases, we assume that the SN is at GC and take three different values of $m_\chi$ indicated by different colors. 

Fig.~\ref{fig:event_per_cos_theta} demonstrates that the arrival BDM are generally concentrated within a small opening angle up to $\theta_{\rm max}$ indicated by the vertical dashed lines. 
The value of $\theta_{\rm max}$ is once again independent of the underlying particle physics model as it is purely determined by the kinematics. 
Since here we integrate over a range of $T_\chi$, $\theta_{\rm max}$ corresponds to the largest $\theta^*$ for $T_{\chi,{\rm min}}$ [see the discussion above Eq.~\eqref{eq:tvan}]. 
Obviously, larger $m_\chi$ give rise to larger $\theta_{\rm max}$, since for a fixed $T_\chi$, the corresponding maximal scattering angle $\psi_{\rm max}$ is larger [see Eq.~\eqref{eq:psimax} in Appendix~\ref{appx:model_kinematics}], reaching $\theta_{\rm max}=0.18\pi$ for $m_\chi= 1$~MeV.

\begin{figure*}
\begin{centering}
\includegraphics[width=0.8\textwidth]{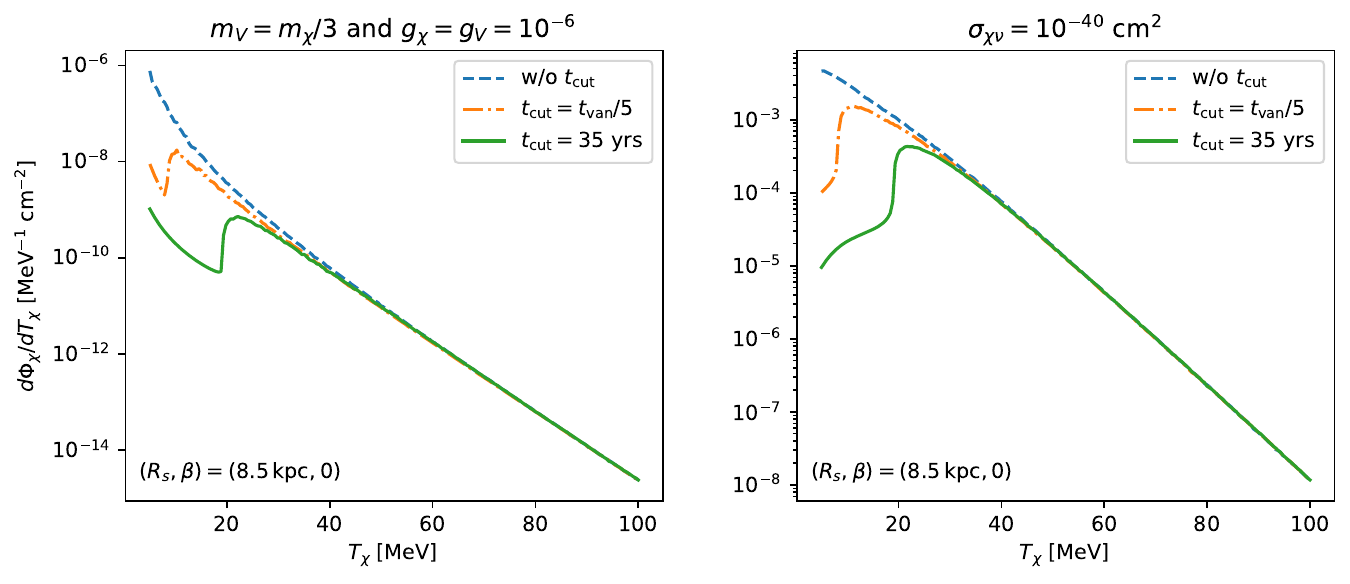}
\end{centering}
\caption{Comparison of the time-integrated SN$\nu$ BDM spectrum between $U(1)_{L_\mu-L_\tau}$ (left) and model-agnostic cross sections (right) with $m_\chi = 1$ MeV.
Blue-dashed line indicates the flux is integrated over time up to $t_{\rm van}$.
The truncation times $t_{\rm cut}=t_{\rm van}/5$ and 35\,yrs are
imposed when $t_{\rm van}>t_{\rm cut}$ for orange-dot-dashed and
green-solid lines, respectively.
}
\label{fig:flux_v_Tx}
\end{figure*}

Comparing the angular distributions obtained with the $U(1)_{L_\mu-L_\tau}$ model to the model-agnostic scenario, one sees that $d\Phi_{\chi}/d\theta$ decreases monotonically in the model-agnostic case, but arises at larger $\theta$ with the $U(1)_{L_\mu-L_\tau}$ model for larger $m_\chi$. 
The reason is as follows. 
For the model-agnostic scenario, the differential cross section $d\sigma_{\chi\nu}/d\cos\psi$ is smaller for larger scattering angle $\psi$ in the lab frame (see e.g., the angular distribution $f_\chi$ as shown in Ref.~\cite{Lin:2022dbl}), which leads to monotonically decreasing  $d\Phi_{\chi}/d\theta$. 
However, for the $U(1)_{L_\mu-L_\tau}$ model, the increase of $d\Phi_{\chi}/d\theta$ at larger $\theta$ is related to the enhanced cross section $d\sigma_{\chi\nu}/d\cos\psi$ at larger $\psi$ as the needed $E_\nu$ is larger. 
Thus, if the detailed angular distribution of SN$\nu$ BDM flux can be reconstructed from the detected events, one would be able to determine the nature of DM-neutrino interaction, in addition to the method using the temporal shape of BDM flux as discussed in Sec.~\ref{sec:res_model}\footnote{For a single BDM-$e$ scattering in water Cherenkov detector such as Super-K, the angular resolution may be as good as $\sim 20\degree$~\cite{Super-Kamiokande:2016yck}. 
However, if large number of events can be accumulated, then better angular resolution due to accumulated statistics may be achieved \cite{Davis:2016hil}.}

\subsection{Time-integrated BDM spectrum}
\label{sec:res_energy}

Finally, it is also of interest to look at the expected energy spectrum of the SN$\nu$ BDM. 
For this purpose, we show in Fig.~\ref{fig:flux_v_Tx} $d\Phi_\chi(T_\chi)/dT_\chi$ by integrating Eq.~\eqref{eq:BDM_flux} over different time duration for cases $m_\chi=1$~MeV.  
The upper time-integration limits are chosen to be $t_{\rm van}=1354\,$yrs (blue-dashed, denoted as w/o $t_{\rm cut}$), $t_{\rm cut}=t_{\rm van}/5$ (orange-dash-dotted), and $t_{\rm cut}=35$~yrs (green-solid), respectively. 
Left and right panels indicate the $U(1)_{L_\mu-L_\tau}$ and model-agnostic cross sections, respectively.

Without imposing any $t_{\rm cut}$, all BDM fluxes arriving at different times at $t<t_{\rm van}$ can contribute to the time-integrated spectrum. 
For both models, $d\Phi_\chi(T_\chi)/dT_\chi$ decreases with $T_\chi$, with slightly different shapes. 
When imposing different values of $t_{\rm cut}$, the lower $T_\chi$ part of the spectrum are suppressed, because a majority part of them only arrive at later times after $t_{\rm cut}$.
The detailed shape of the BDM spectrum at smaller $T_\chi$ once again depends on the specific underlying interaction model, as well as the SN location, as both affect the temporal evolution of the BDM (see Secs.~\ref{sec:res_SN_loc} and~\ref{sec:res_model}). 

\begin{figure*}
\begin{centering}
\includegraphics[width=0.8\textwidth]{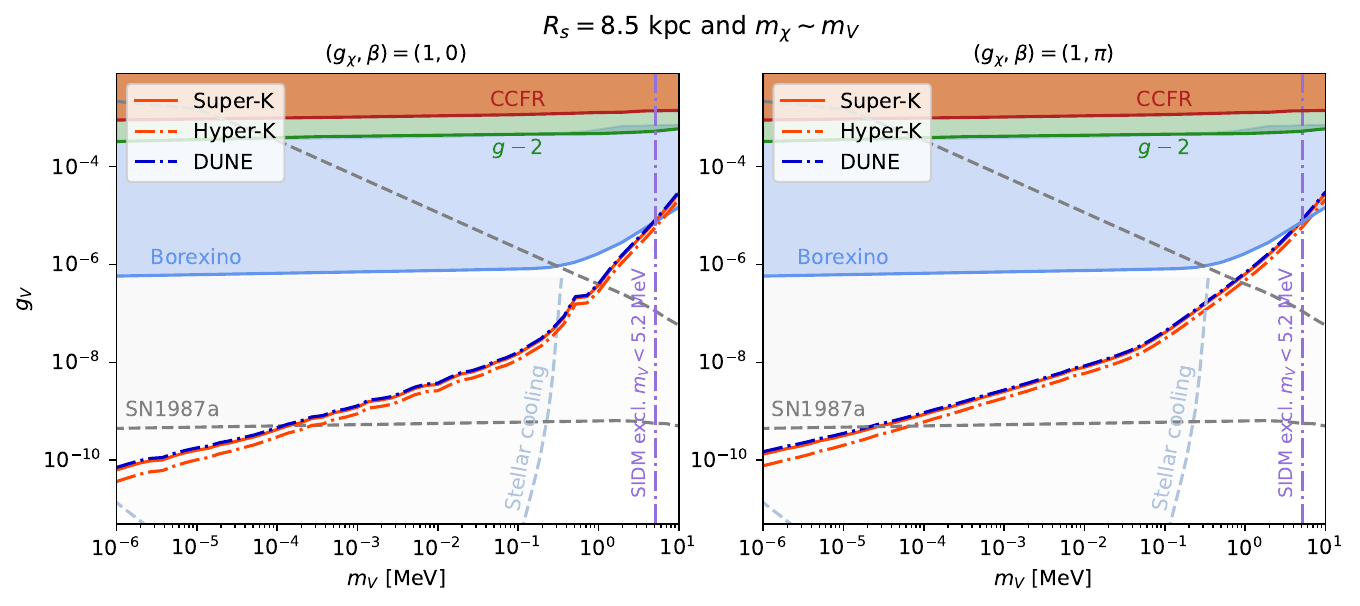}
\includegraphics[width=0.8\textwidth]{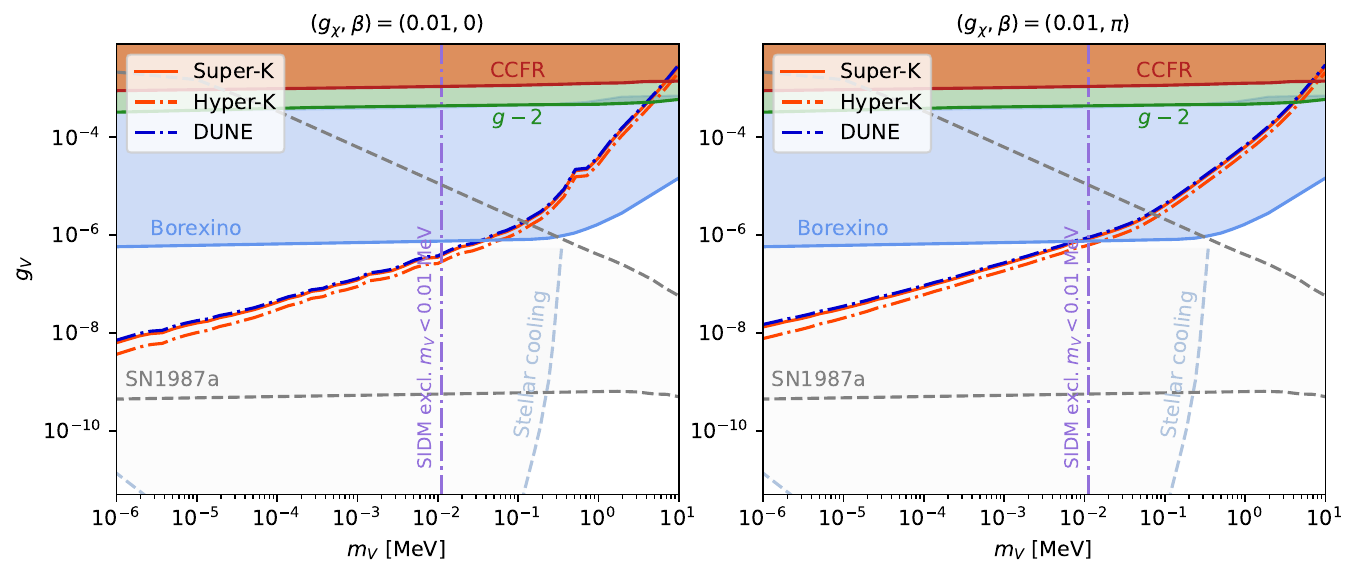}
\end{centering}
\caption{Projected sensitivities on $g_V$ as functions of $m_V$ for Super-K, Hyper-K and DUNE with the parameter choice of $\varepsilon=-g_V$ and $R_s=8.5$ kpc.
Upper and lower panels correspond to results with $g_\chi=1$ and $0.01$, respectively.
Left and right panels show results obtained by taking   $\beta=0$ and $\pi$, respectively. 
To obtain these plots, thermal relic constraint is imposed, which results in $m_\chi\sim m_V$ (see Appendix~\ref{appx:constraints_cosmos_sidm}). 
Together with the constraint from DM self-interaction \cite{Randall:2008ppe,Walker:2011zu,Boylan-Kolchin:2011lmk,Boylan-Kolchin:2011qkt,Elbert:2014bma,Tulin:2017ara,Adhikari:2022sbh}, which depends on $g_\chi$, the detection sensitivities exclude certain regions of $m_V$.
Color shaded regions are the existing bounds \cite{Amaral:2020tga,Pospelov:2008zw,Escudero:2019gzq,PhysRevLett.66.3117,Altmannshofer:2014pba,Croon:2020lrf}.
}
\label{fig:sensitivity_gx_equal_eps}
\end{figure*}
\section{Sensitivity analysis}\label{sec:sensitivity}

\begin{figure*}
\begin{centering}
\includegraphics[width=0.8\textwidth]{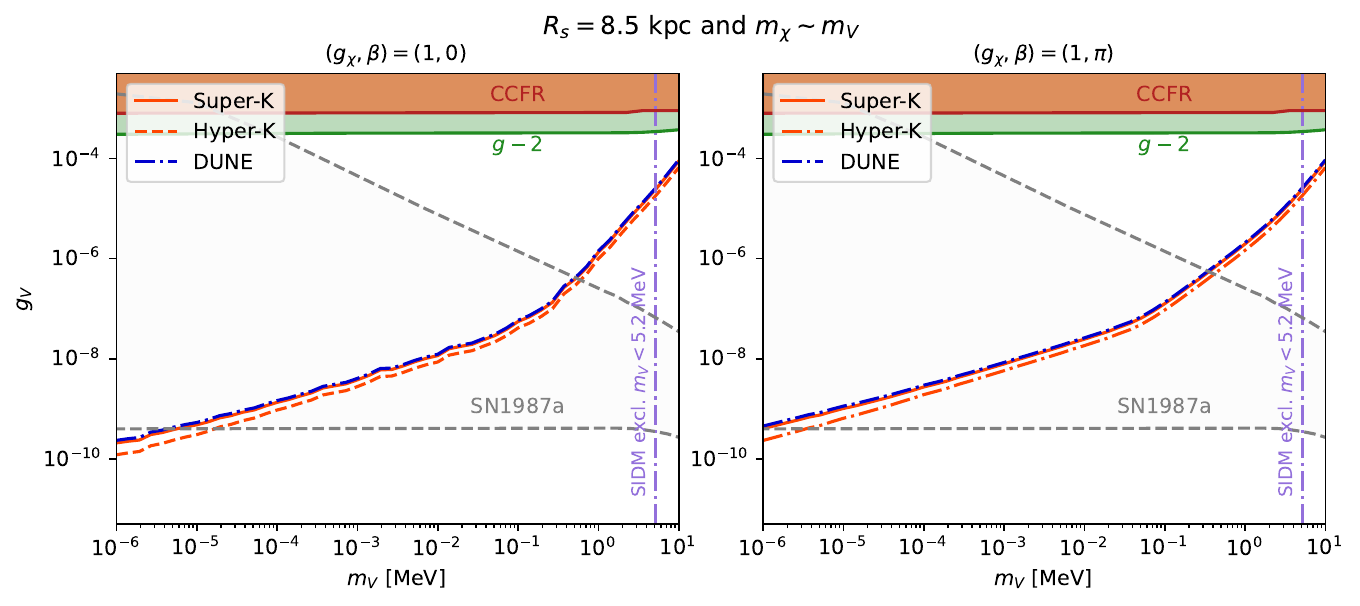}
\includegraphics[width=0.8\textwidth]{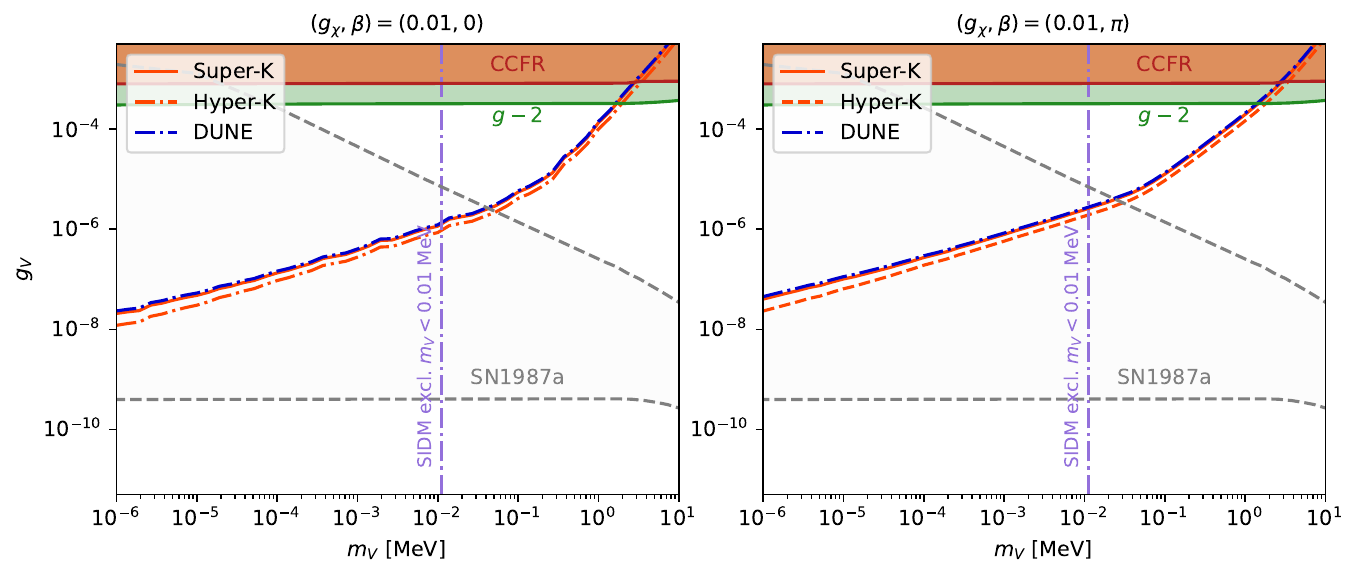}
\end{centering}
\caption{
Projected sensitivities on $g_V$ as functions of $m_V$ for Super-K, Hyper-K and DUNE with zero kinetic mixing $\varepsilon=0$. 
All other parameters are the same as those used for  Fig.~\ref{fig:sensitivity_gx_equal_eps}. Note that the bounds from Borexino and stellar cooling
are not applicable in this case.}
\label{fig:sensitivity_gx_noeps}
\end{figure*}

To analyze sensitivity on model parameters, 
we evaluate the required event number for signal, $N_s$, via
\begin{equation}
    2.0 = \frac{N_s}{\sqrt{N_s+N_b}}\label{eq:sensitivity}
\end{equation}
where 2.0 implies the $2\sigma$ detection significance and $N_b$ is the background event number.
We can match $N_s$ with Eq.~\eqref{eq:dPhi/dtheta} by integrating it over $\theta$,
\begin{equation}
    N_s = \int_0^{\pi/2} d\theta \frac{d\Phi_\chi}{d\theta}
\end{equation}
where the ranges for $T_\chi$ is $(5,100)$ MeV and $t$ is $(t_0,t_{\rm exp})$ as discussed in the last section.
A table of $N_s$ for different model parameters is listed in
Appendix \ref{appx:BDM_numbers}.
We adopt $N_b\sim \mathcal{B}M_T t_{\rm exp}$ where $M_T$ is the detector fiducial mass and $t_{\rm exp}$ the exposure time.
We have $M_T=22.2$ kton for Super-K, 222 kton for Hyper-K and 17 kton for DUNE.
The associated $N_e$ are $7.34\times 10^{33}$, $7.34\times 10^{34}$ and $4.58\times 10^{33}$, respectively.
The factor $\mathcal{B}$ has the unit per kton per year and varies with the $T_\chi$ range of interest.
As shown in Ref.~\cite{Super-Kamiokande:2016yck}, $\mathcal{B}=526$ for $T_{\chi,{\rm min}}=5\,{\rm MeV}$ in water Cherenkov detector.
For DUNE detector, $\mathcal{B}=427$ by a rescaling via the ratio of total electron number per kton between water and liquid argon.
The major contribution to the estimated background above is originated from solar neutrinos between 5\,MeV to 25\,MeV. 
In addition, muon spallation is also known to be an important background for solar neutrino measurement in the energy range between 6 MeV to 18 MeV~\cite{Super-Kamiokande:2016yck}. 
However, multiple cuts can be applied in Super-K to remove the spallation background \cite{Li:2014sea,Super-Kamiokande:2016yck}. 
Thus, we assume that such cuts can be similarly applied in Hyper-K \cite{Hyper-Kamiokande:2018ofw} and DUNE \cite{Zhu:2018rwc} as well and therefore neglect the spallation background in our analysis. 
For background events from the radon radioactivity, they only contribute to the energy range below 5 MeV \cite{Nakano:2017hpu}, which is not relevant in this study.
For energies higher than 25 MeV, atmospheric neutrinos \cite{Battistoni:2005pd,Dutta:2019oaj} are the primary backgrounds. 
We discuss another choice for $T_{\chi,{\rm min}}=25$ MeV in Appendix~\ref{appx:Txmin_25MeV}.
Although this choice avoids solar neutrinos in $\mathcal{B}$, the BDM events are also significantly suppressed\footnote{Note that here we do not consider other possible background contributions that are generally considered in the solar and atmospheric neutrino experiments. Including those might slightly affect our derived sensitivities but the overall effect are estimated to be small.}.

As shown by $\mathcal{L}_X$, Eq.~\eqref{eq:Lagrangian}, the free parameters to be constrained are $g_V$, $g_\chi$, $\varepsilon$, $m_\chi$ and $m_V$.
To proceed, we first apply the thermal relic constraint which states that the DM annihilation cross section $\langle \sigma v\rangle$ is fixed by the canonical value $6\times 10^{-26}~{\rm cm^3\,s^{-1}}$
for Dirac-fermion DM. This restricts $m_V\sim m_\chi$ assuming that $\chi\bar{\chi}\to 2V$ is 
the dominant annihilation channel.
Given the similarity between the dark charge $g_\chi$ and the electric charge $e$, we take $g_\chi=0.01$ and $1$, which guarantees the validity of perturbation expansions in our study. 
With $g_\chi$ fixed, the self-interacting DM (SIDM) constraint from astrophysical observations \cite{Randall:2008ppe,Walker:2011zu,Boylan-Kolchin:2011lmk,Boylan-Kolchin:2011qkt,Elbert:2014bma,Tulin:2017ara,Adhikari:2022sbh},
sets a lower bound for $m_V$.
See Appendix~\ref{appx:constraints_cosmos_sidm} for relevant discussions.
We study cases with and without the presence of the kinetic mixing.
Without losing the generality, we take $\varepsilon=-g_V$ when kinetic mixing is non-vanishing.
This naturally enables DM scattering with electrons in the detector.
However, in the absence of $\varepsilon$, the DM-$e$ interaction is only possible through
$\mu/\tau$ loop shown in Fig.~\ref{fig:mutau_loop}.
It induces a non-zero coupling $\varepsilon^\prime$ between $V$ and SM photon as given by Eq.~\eqref{eq:induced_eps}.

\subsection{Case study for $\varepsilon=-g_V$}

Fig.~\ref{fig:sensitivity_gx_equal_eps} shows the sensitivities on $g_V$ as functions of $m_V$ in Super-K, Hyper-K and DUNE. 
The results are presented together with the existing
bounds from Borexino~\cite{Amaral:2020tga}, muon $g-2$~\cite{Pospelov:2008zw,Escudero:2019gzq}, CCFR\footnote{This name stands for Columbia-Chicago-Fermilab-Rochester experiment that we borrowed from Ref.~\cite{Croon:2020lrf}. It measures $\mu^-\mu^+$ pair production due to
$\nu$ scattering with nucleus in the Coulomb field.
This is a rare process and known as the neutrino trident.} \cite{PhysRevLett.66.3117,Altmannshofer:2014pba}, stellar cooling~\cite{An:2013yfc,Hardy:2016kme} and
SN1987A~\cite{Croon:2020lrf}.
One could also present the sensitivities/constraints with $g_V-m_\chi$ parameter regions by taking $m_\chi\sim m_V$ as implied by the thermal relic constraint.  
Note that the two astrophysical constraints are plotted with dashed lines in Fig.~\ref{fig:sensitivity_gx_equal_eps} as they were derived by considering the excess cooling due to the presence of light dark mediator $V$ only. 
When the entire light dark sector is taking into consideration, their self-trapping may help evade these bounds~\cite{Sung:2021swd}.
Upper and lower panels are results obtained by taking $g_\chi = 1$ and $0.01$, respectively. Left and right panels correspond to the choice of $\beta=0$ and $\pi$, respectively.
The SIDM constraint disfavors $m_V$ to the left of the purple dot-dashed line shown in Fig.~\ref{fig:sensitivity_gx_equal_eps}.
When $g_\chi$ is smaller, the SIDM constraint permits a lighter $m_V$ as $\sigma_{\chi\chi}\propto g_\chi^4m_V^{-2}$ for $m_\chi\sim m_V$, Eq.~\eqref{eq:sigxx}.

\begin{figure*}
\begin{centering}
\includegraphics[width=0.8\textwidth]{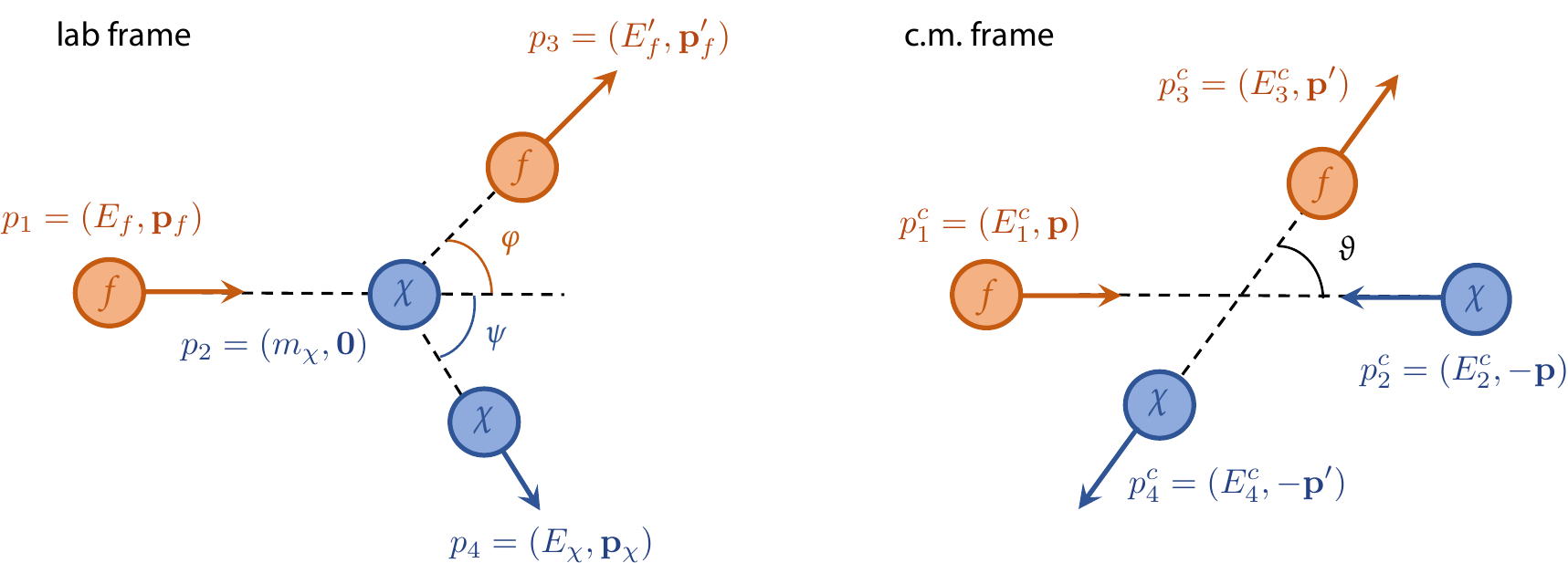}
\end{centering}
\caption{\label{fig:geometry_scatter}
Halo DM $\chi$ scattered off by fermion $f$ in the lab (left) and c.m.~(right) frames.}
\end{figure*}

Since $N_s$ is proportional to $g_\chi^4 g_V^2\varepsilon^2 N_e$ with $\varepsilon=-g_V$, the sensitivity on $g_V$ for a fixed $m_V$ scales as $g_{\chi}^{-1}$.
Furthermore, a larger number of target electrons naturally leads to a better sensitivity.
This explains why Hyper-K is more sensitive than the other two experiments. The sensitivity for $\beta=\pi$ is weaker than that for $\beta=0$, since SN$\nu$'s in the former case propagate through those galactic halo regions with much smaller DM densities.  
With non-zero kinetic mixing, the sensitivity region for SN$\nu$ BDM is tightly constrained by 
the existing bounds.
For $g_\chi=1$, a small viable window  
$0.3~{\rm MeV}<m_V<10\,{\rm MeV}$ appears, which   
is however disfavored once SIDM constraint is imposed.
\subsection{Case study for $\varepsilon=0$}

In Fig.~\ref{fig:sensitivity_gx_noeps}, we present the sensitivities with zero kinetic mixing.
Without $\varepsilon$, the DM-$e$ interaction 
can only occur through the $\mu/\tau$-loop (see Fig.~\ref{fig:mutau_loop}),
with the induced kinetic mixing parameter $\varepsilon^\prime$.
In general, $\varepsilon^\prime$ varies with the momentum transfer $q$ as shown by Eq.~\eqref{eq:induced_eps}.
For our interested $T_\chi$ range, 
the approximation $\varepsilon^\prime \approx -g_V/70$ holds.
With a much suppressed DM-$e$ interaction strength, 
a much larger $g_V$ is needed 
to attain the required $N_s$ for the detection sensitivity.
This explains why the results shown in Fig.~\ref{fig:sensitivity_gx_noeps} are weaker than those in Fig.~\ref{fig:sensitivity_gx_equal_eps} by a factor of $\sqrt[4]{4900}\approx 8.37$ since $N_s \propto g_V^2 \varepsilon^{\prime 2}=g_V^4/4900$.

In the absence of kinetic mixing, bounds like Borexino and stellar cooling are not applicable.
This leaves more window for SN$\nu$ BDM to be probed.
For $g_\chi=1$ (upper panel of Fig.~\ref{fig:sensitivity_gx_noeps}), 
SN$\nu$ BDM can probe the mass range $m_V\gtrsim 5$ MeV which is not constrained by SIDM.
For $g_\chi=0.01$, SIDM constraint only disfavors $m_V< 0.01$ MeV. 
The sensitive region by SN$\nu$ BDM is then expanded to $10~{\rm keV}<m_V<\mathcal{O}(1)~{\rm MeV}$ and $\mathcal{O}(10^{-5})<g_V<\mathcal{O}(10^{-3})$.

\section{Summary}\label{sec:summary}

The BDM accelerated by energetic cosmic particles have been demonstrated in recent years as an important component that help extend the sensitivity of existing and future DM and/or neutrino experiments to search for light DM with $m_\chi<\mathcal{O}(100)\,{\rm MeV}$. 
SN$\nu$~BDM not only shares this advantage but also acquires additional TOF information, which may allow direct inference of $m_\chi$ with the method proposed in Ref.~\cite{Lin:2022dbl}. 
In this paper, we built upon Ref.~\cite{Lin:2022dbl} 
and considered generalized scenarios that a SN explosion can happen at an arbitrary location in the galaxy. 
We found that the time profile of the SN$\nu$~BDM can depend on the SN location. 
For a SN off the GC, although the SN$\nu$~BDM flux for a given $T_\chi$ does not contain a dominant peak, it still possesses a distinct sharp rise at a specific time and vanishes at a later time, both of which only depend on the distance to SN and the DM mass. 
As a result, 
if the SN occurring time and location can be independently measured with its multimessenger signals,
a detailed measurement of the temporal $T_\chi$ dependent flux of SN$\nu$~BDM can be used as a TOF experiment to infer the DM mass.

Moreover, we have also gone beyond the simplified model-independent assumption made in Ref.~\cite{Lin:2022dbl} and considered SN$\nu$ BDM based on the well-studied 
$U(1)_{L_\mu-L_\tau}$ model where the gauge boson $Z'$ gives rise to  non-vanishing cross sections, $\sigma_{\chi\nu}$ and $\sigma_{\chi e}$.
We found that not only the exact temporal but also the expected angular distribution as well the energy spectrum of the SN$\nu$~BDM flux contain features related to the underlying particle physics model assumption. 
This indicates that a precisely measurement of the SN$\nu$~BDM fluxes can in principle be used to discern the type of interaction connecting the SM and the dark sector.



By considering the existing and upcoming large scale neutrino experiments including Super-K, Hyper-K, and DUNE, we derived their projected SN$\nu$ BDM sensitivities on $g_V$ versus $m_V$ for the $U(1)_{L_\mu-L_\tau}$ model, and compare them to other existing bounds, assuming that the DM are produced as thermal relic. 
With non-zero kinetic mixing, the parameter space that can be probed by SN$\nu$ BDM is mostly constrained by terrestrial experiments together with the limit derived for SIDM. 
However, for cases with zero kinetic mixing, SN$\nu$ BDM can probe a significant part of the yet-constrained parameter space for $m_V$ from $\mathcal{O}({\rm keV})$ to $\mathcal{O}({\rm MeV})$.


While we have only considered exclusively the DM leptonic interaction throughout this work, we note that 
the detection of SN$\nu$ BDM 
through other channels are possible and should be explored. 
Taking DUNE for instance, argon deexcitation provides a much cleaner detection channel for
low energy event, $T_\chi \sim \mathcal{O}({\rm MeV})$.
This signal is possible in phenomenological models like $B-L$ \cite{Croon:2020lrf} or models that directly involve DM-quark couplings. 
In addition, for certain DM mass range, the SN$\nu$ BDM energy spectra may peak at low kinetic energy, which may result in better sensitivities when considering current or upcoming DM experiments with very low energy threshold such as XENONnT.
A more comprehensive study of DM-SM couplings through the effective Lagrangian and the investigations of signatures in different detectors will be carried out in future work. 

The SN$\nu$ BDM afterglow is an intriguing astrophysical phenomenon that ties to physics beyond SM.  
The features from the time-evolving BDM flux open a new avenue for probing light DM. We also anticipate the general framework surveyed in this paper can facilitate further developments in new detection methods.

\begin{acknowledgments}
The authors thank Cheng-Wei Chiang,	Bhaskar Dutta, Kate Scholberg for useful discussions, and in particular Alec Habig for insightful exchanges and for providing feedback to our manuscript.  
Y.-H.~L.~acknowledges the support from National Science and Technology Council (NSTC), the Ministry of Education under Project No.~NTU-112L104022, and the Physics Division, National Center for Theoretical Sciences of Taiwan (PD-NCTS). G.-L.~L. acknowledges the support from NSTC under Grant No.~111-2112-M-A49-027. He also likes to thank the Institute of Physics, Academia Sinica for the warm hospitality when this work was initiated. 
H.~T.-K.~W.~acknowledges supports from the NSTC under Grant No.~110-2112-M-001-029-MY3
M.-R.~W.~acknowledges supports from the NSTC under Grant No.~110-2112-M-001-050 and 111-2628-M-001-003-MY4, the Academia Sinica under Project No.~AS-CDA-109-M11, and PD-NCTS.
We also acknowledge the computing resources provided by the Academia Sinica Grid-computing Center.
\end{acknowledgments}

\appendix

\section{The scattering cross sections}\label{appx:model_kinematics}



\subsection{Lorentz invariant differential cross section}

The scattering process for two particles $f$ and $\chi$ in c.m.~frame is shown in the right panel of Fig.~\ref{fig:geometry_scatter},
and the differential cross section is given by
\begin{equation}
    \frac{d\sigma_{\chi f}}{d\Omega^*} = \frac{1}{64\pi^2 s_M}\frac{|\mathbf{p}^\prime|}{|\mathbf{p}| }|\mathcal{M}|^2\label{eq:diff_crox_xf_cm} 
\end{equation}
where $|\mathcal{M}|^2$ is the scattering amplitude, $d\Omega^*=d\cos\vartheta d\phi^*$ and $\phi^*$ the azimuthal angle in c.m.~frame.
Assuming elastic scattering, we can apply $|\mathbf{p}|=|\mathbf{p}^\prime|$ and take
\begin{equation}
    u_M=(p_1-p_4)^2=m_f^2 + m_\chi^2 - 2(E_1^c E_4^c + |\mathbf{p}||\mathbf{p}^\prime| \cos\vartheta)
\end{equation}
hence 
\begin{alignat}{1}
    du_M &= -2|\mathbf{p}||\mathbf{p}^\prime|  d\cos\vartheta \nonumber \\
    & \to d\Omega^* = d\cos\vartheta d\phi^* = -\frac{du_M d\phi^*}{2|\mathbf{p}||\mathbf{p}^\prime| }.
\end{alignat}
We thus recast Eq.~\eqref{eq:diff_crox_xf_cm} into frame-independent differential cross section
\begin{equation}
    \frac{d\sigma_{\chi f}}{du_M} = -\frac{|\mathcal{M}|^2}{64\pi s_M |\mathbf{p}|^2}.\label{eq:fi_dsig_xf}
\end{equation}
The c.m.~frame azimuthal angle $\phi^*$ has been integrated out and the momentum-squared is
\begin{equation}
    |\mathbf{p}|^2=\frac{1}{4s_M}[s_M-(m_f+m_\chi)^2][s_M-(m_f-m_\chi)^2]\label{eq:p_square}
\end{equation}
where
\begin{equation}
    s_M = m_f^2+m_\chi^2 + 2E_f m_\chi
\end{equation}
and $m_{\chi,f}$  are the associated masses.
Therefore, we can boost Eq.~\eqref{eq:fi_dsig_xf} into any frame of interest.

\subsection{The differential $\nu\chi$ scattering cross section in lab frame}\label{appx:diff_xv_crox}

\begin{figure}
\begin{centering}
\includegraphics[width=0.9\columnwidth]{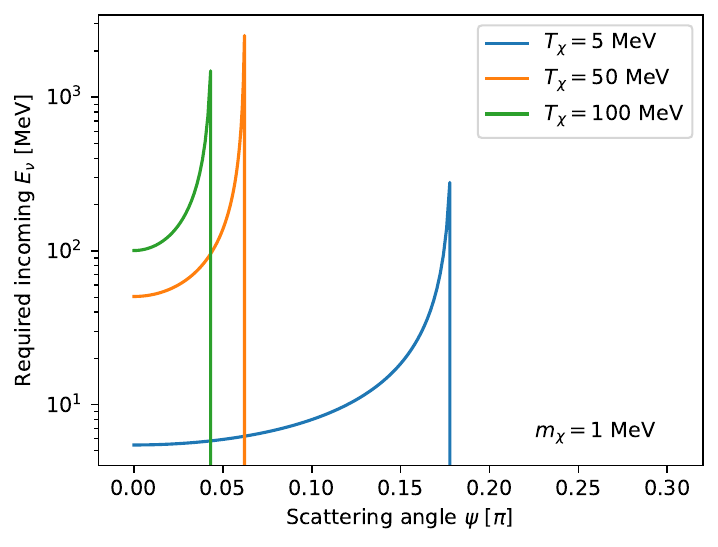}
\end{centering}
\caption{\label{fig:required_Ev}
The required $E_\nu$ versus scattering angle $\psi$ for different $T_\chi$ with $m_\chi=1$ MeV.
}
\end{figure}

\begin{figure}
\begin{centering}
\includegraphics[width=0.9\columnwidth]{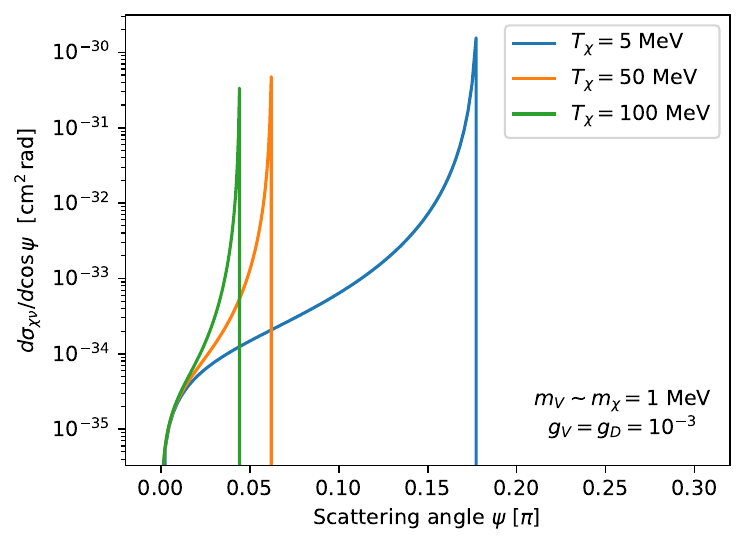}
\end{centering}
\caption{\label{fig:diffcrox_xnu}
The differential DM-$\nu$ cross section versus scattering angle $\psi$.
Model parameters are shown on the plot.
}
\end{figure}

We now boost Eq.~\eqref{eq:diff_crox_xf_cm} into lab-frame-dependent differential cross section and replace $f$ by $\nu$, 
\begin{equation}
    \frac{d\sigma_{\chi\nu}}{d\Omega} = \frac{d\sigma_{\chi\nu}}{du_M}\frac{du_M}{d\Omega}
    =\frac{1}{2\pi}\frac{du_M}{d\cos\psi}\frac{d\sigma_{\chi\nu}}{du_M}\label{eq:dsigXVdOmega}
\end{equation}
where the azimuthal angle is already integrated out.
Assuming $\nu$ is massless and letting  $|\mathbf{p}_\chi|=p_\chi$, the four momenta $p_{1,2,3,4}$ are (left panel of Fig.~\ref{fig:geometry_scatter})
\begin{subequations}
\begin{align}
    p_1 &= (E_\nu,0,0,E_\nu),\\
    p_2 &= (m_\chi,0,0,0),\\
    p_3 &= (E_\nu^\prime,E_\nu^\prime \sin\varphi,0,E_\nu^\prime \cos\varphi),\\
    p_4 &= (E_\chi,-p_\chi \sin\psi,0,p_\chi \cos\psi).
\end{align}
\end{subequations}
We then have 
\begin{subequations}
\begin{alignat}{1}
    u_M &= (p_1-p_4)^2 = m_\chi^2 - 2(E_\nu E_\chi-E_\nu p_\chi \cos\psi),\label{eq:u_p14}\\
      &= (p_2-p_3)^2 = 
       2m_\chi (E_\chi-E_\nu) - m_\chi^2.\label{eq:u_p23} 
\end{alignat}
\end{subequations}
Let $E_\nu^\prime = E_\nu - E_\chi + m_\chi$ in the last line and having $E_\chi = m_\chi + T_\chi$. By equating Eqs.~\eqref{eq:u_p14} and \eqref{eq:u_p23} we get
\begin{equation}
    E_\nu = \frac{m_\chi(m_\chi-E_\chi)}{E_\chi-m_\chi-p_\chi\cos\psi}=-\frac{m_\chi T_\chi}{T_\chi -p_\chi\cos\psi}.\label{eq:Ev}
\end{equation}
and
\begin{equation}
    \frac{dE_\nu}{dT_\chi} = \frac{m_\chi^2 T_\chi \cos\psi}{p_\chi(T_\chi - p_\chi\cos\psi)^2}.\label{eq:dEv/dTx}
\end{equation}
Eq.~\eqref{eq:Ev} can be used to determine the required $E_\nu$ with specified $T_\chi$ and $\psi$.
Note that Eq.~\eqref{eq:Ev} becomes negative as $T_\chi < p_\chi\cos\psi$ and is unphysical.
The range of $\psi$ is constrained by
\begin{equation}\label{eq:psimax}
    0<\psi<\psi_{\rm max}=\cos^{-1}\left(\frac{T_\chi}{p_\chi}\right)
\end{equation}
See Fig.~\ref{fig:required_Ev} for numerical computation. 
With Eq.~\eqref{eq:u_p14}, we can rewrite Eq.~\eqref{eq:diff_crox_xf_cm} in
terms of $\psi$ by
\begin{equation}
    \frac{d\sigma_{\chi\nu}}{d\cos\psi}=\frac{1}{32\pi}\sqrt{\frac{1}{m_\chi^3}\left(\frac{1}{m_\chi}+\frac{2}{T_\chi}\right)}|\mathcal{M}|^2,\label{eq:diff_sigxv_psi}
\end{equation}
where the amplitude $|\mathcal{M}|^2$ is given by Eq.~\eqref{eq:amplitude}
with $m_{1,2}$ replaced by $m_{\chi,\nu}$.
Combining Eq.~\eqref{eq:Ev}, we display the numerical results for Eq.~\eqref{eq:diff_sigxv_psi}
in Fig.~\ref{fig:diffcrox_xnu}. 
When $\psi$ approaches $\psi_{\rm max}$, the required $E_\nu$ increases rapidly,
thus the differential cross section is enhanced accordingly.

As discussed in the main text, although $d\sigma_{\chi\nu}/d\cos\psi$ increases for
large $\psi$, typically corresponding to large open angle $\theta$, see Fig.~\ref{fig:geometry_scheme}, the BDM flux regarding on large $\theta$ will be 
suppressed eventually due to the exponential suppression in $f_{\nu_i}$ in Eq.~\eqref{eq:dnv/dEv}.

\subsection{The total $\chi e$ scattering cross section}

To obtain the total $\chi e$ cross section, we apply Eq.~\eqref{eq:fi_dsig_xf} and convert $u_M$ to $t_M$ via $du_M=-dt_M$,
\begin{equation}
     \frac{d\sigma_{\chi e}}{dt_M}=\frac{|\mathcal{M}|^2}{64\pi s_M |\mathbf{p}|^2}.
\end{equation}
The total cross section can be obtained by integrating over $t$,
\begin{equation}
    \sigma_{\chi e} =\int_{t_M^-}^{t_M^+} \frac{d\sigma_{\chi e}}{dt_M} dt_M
\end{equation}
with 
\begin{equation}
    t_M^\pm = m_\chi^2+m_e^2-2(E_\chi^c E_\chi^{c\prime} \mp |\mathbf{p}||\mathbf{p^\prime}|).
\end{equation}
and
\begin{subequations}
\begin{gather}
    E_\chi^c = E_\chi^{c\prime} = \frac{1}{2\sqrt{s_M}}(s_M +m_\chi^2-m_e^2),\\
    s_M = m_\chi^2+m_e^2+2(m_\chi+T_\chi) m_e,\\
    |\mathbf{p}| = |\mathbf{p^\prime}| \to |\mathbf{p}||\mathbf{p^\prime}| = {\rm Eq.~}\eqref{eq:p_square}.\nonumber
\end{gather}
\end{subequations}

\section{Geometrical relations}\label{appx:geometry}

\begin{figure*}
\begin{centering}
\includegraphics[width=0.8\textwidth]{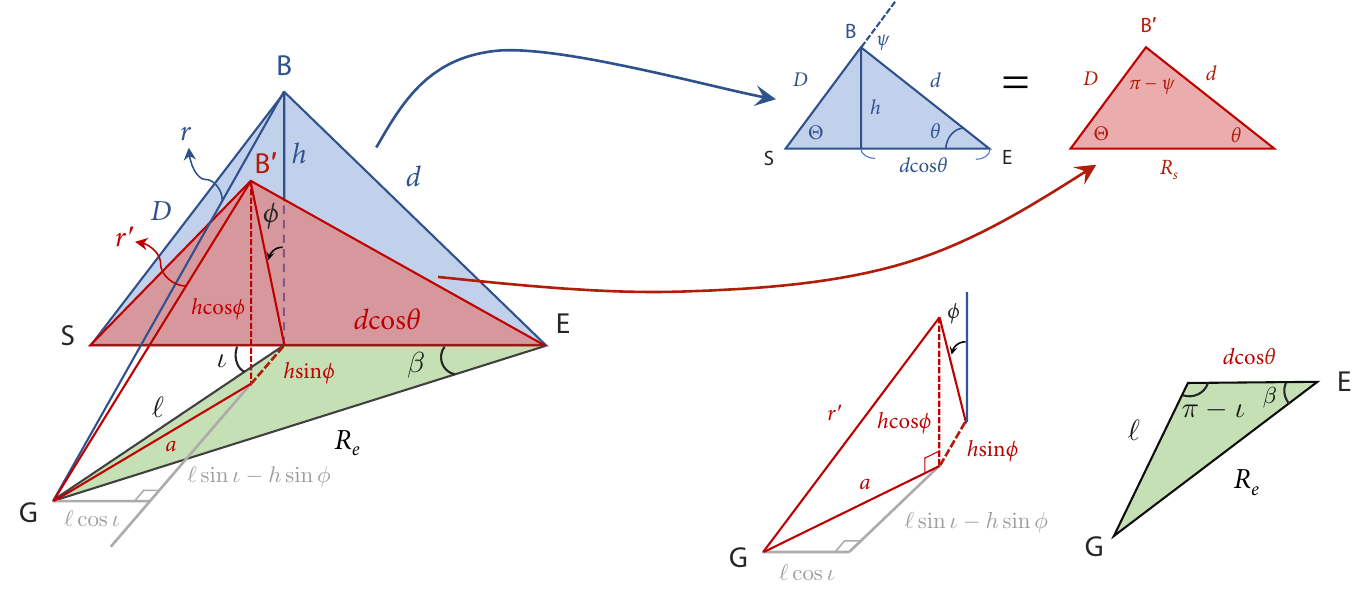}
\end{centering}
\caption{\label{fig:geometry_anatomy}
The 3D diagram that depicts the geometical relations between $\mathsf{G}$, $\mathsf{S}$, $\mathsf{E}$ and $\mathsf{B}$.}
\end{figure*}

In this appendix, we show in detail how one obtains the geometrical relations for calculating SN$\nu$ BDM on Earth, see Figs.~\ref{fig:geometry_topView} and \ref{fig:geometry_scheme}. The derivations and results presented here are coordinate independent. One can apply these results to, for instance, Eq.~\eqref{eq:BDM_flux} and employ a particular coordinate system for numerical calculations.

In principle, once $\triangle \mathsf{SBE}$ is determined, the related geometries are similar for both SN at-GC and off-GC cases.
The only difference is the DM number density $n_\chi$ at the boosted point $\mathsf{B}$. To determine $n_\chi$ at $\mathsf{B}$, one needs to know the distance from it to $\mathsf{G}$.
In terms of at-GC, $\overline{\mathsf{BG}}$
is independent of the azimuthal angle $\phi$ relative to the SN-Earth-axis ($\mathsf{SE}$). 
This does not hold for off-GC case.
Let $\mathsf{B}$ corresponds to $\phi=0$ and $\mathsf{B}^\prime$ to $\phi\neq 0$,  Fig.~\ref{fig:geometry_anatomy}.
The distances from $\mathsf{B}$ and $\mathsf{B}^\prime$ to GC are $r$ and $r^\prime$ where $r\neq r^\prime$ in general.
The goal is to determine $r^\prime$ in order to compute $n_\chi$ at $\mathsf{B}^\prime$.

In Fig.~\ref{fig:geometry_anatomy}, $\triangle \mathsf{SBE}$ (blue) and $\triangle \mathsf{SB^\prime E}$ (red)
are congruent. But the red one is placed at non-zero $\phi$
along $\mathsf{SE}$-axis.
Let $\theta$ be the open angle and $\psi$ the BDM scattering angle at boosted point.
The identity
\begin{equation}
    D^2 = d^2 + R_s^2-2dR_s\cos\theta.\label{eq:D2}
\end{equation}
holds.
Given $t=t^\prime-t_\nu$ and having SN$\nu$ moving with the light speed $c$, we have
\begin{equation}
    \frac{D}{c} + \frac{d}{v_\chi} = t^\prime \to D + \frac{d}{\beta_\chi} = R_s + ct \equiv \zeta\label{eq:timing}
\end{equation}
where $\beta_\chi=v_\chi/c$.
Plugging Eq.~\eqref{eq:D2} into Eq.~\eqref{eq:timing}, we obtain
\begin{alignat}{1}
    d   &=-\frac{\beta_\chi}{1-\beta_\chi^2}\left[\sqrt{(R_s^2-\zeta^2)(1-\beta_\chi^2)+(R_s \beta_\chi\cos\theta-\zeta)^2}\right.\nonumber \\
     &  \quad+ \left. R_s \beta_\chi\cos\theta 
-\zeta  \frac{}{}\right].\label{eq:d}
\end{alignat}
To compute $d\Phi_\chi/d T_\chi$, it takes two inputs: $T_\chi$ and $t^\prime$ (equivalent to $t$ by a constant shift $t_\nu$).
With the aid of Eq.~\eqref{eq:d}, the relevant geometrical relations can be obtained.
Therefore,
\begin{alignat}{1}
r^{\prime2} & =a^{2}+h^{2}\cos^{2}\phi\nonumber \\
 & =\ell^{2}\cos^{2}\iota+(\ell\sin\iota-h\sin\phi)^{2}+h^{2}\cos^{2}\phi,\label{eq:r_prime}
\end{alignat}
and 
\begin{equation}
    \ell^2 = R_e^2+ d^2\cos^2\theta -2R_e d\cos\theta \cos\beta.\label{eq:ell}
\end{equation}
The parameters $R_e$, $d$, $\theta$ and $\beta$ should be specified already.
Through the law of cosine, we have
\begin{equation}
    \cos\iota = \frac{R_e^2-\ell^2-d^2\cos^2\theta}{2\ell d\cos\theta}.\label{eq:cos_iota}
\end{equation}
Putting Eqs.~\eqref{eq:ell} and \eqref{eq:cos_iota} back into Eq.~\eqref{eq:r_prime}, 
$r^\prime$ can be evaluated. Therefore, $n_\chi$ at $\mathsf{B}^\prime$ is also determined.
One can do a cross check that $r^\prime=r$ at $\phi=0$ and
$r^\prime=D$ for at-GC case.

\section{Thermal relic and SIDM constraints}\label{appx:constraints_cosmos_sidm}

\begin{table*}
\begin{centering}
\begin{tabular}{ccccc}
\hline 
 & \multicolumn{2}{c}{$\varepsilon=-g_{V}$} & \multicolumn{2}{c}{$\varepsilon=0$}\tabularnewline
\hline 
$m_{\chi}$ {[}MeV{]} & $\beta=0$ & $\beta=\pi$ & $\beta=0$ & $\beta=\pi$\tabularnewline
\hline 
\hline 
$10^{-6}$ & $2.51\times10^{13}$ & $1.28\times10^{12}$ & $2.17\times10^{11}$ & $1.56\times10^{10}$\tabularnewline
$10^{-5}$ & $6.89\times10^{11}$ & $4.46\times10^{10}$ & $8.79\times10^{9}$ & $4.30\times10^{8}$\tabularnewline
$10^{-4}$ & $2.31\times10^{10}$ & $1.37\times10^{9}$ & $2.32\times10^{8}$ & $1.36\times10^{7}$\tabularnewline
$10^{-3}$ & $8.50\times10^{8}$ & $4.42\times10^{7}$ & $8.99\times10^{6}$ & $4.80\times10^{5}$\tabularnewline
$10^{-2}$ & $3.41\times10^{7}$ & $1.33\times10^{6}$ & $3.01\times10^{5}$ & $1.47\times10^{4}$\tabularnewline
$10^{-1}$ & $1.63\times10^{5}$ & $3.11\times10^{3}$ & $8.97\times10^{2}$ & $33.20$\tabularnewline
1 & $0.42$ & $5.16\times10^{-2}$ & $2.60\times10^{-3}$ & $5.63\times10^{-4}$\tabularnewline
10 & $1.26\times10^{-8}$ & $1.26\times10^{-8}$ & $1.36\times10^{-10}$ & $1.36\times10^{-10}$\tabularnewline
\hline 
\end{tabular}
\par\end{centering}
\caption{\label{tab:Ns} 
Expected SN$\nu$ BDM event number 
$N_{s}$ vs.~$m_{\chi}$ for the different choices of  $R_s=8~{\rm kpc},(g_{V}=10^{-5},g_{\chi}=10^{-2},\varepsilon,\beta)$ for Super-K.
}

\end{table*}

The DM annihilation cross section $\langle \sigma v\rangle$ for the vector type interaction was  derived in Ref.~\cite{Lin:2021hen}.
Depending on $m_V$, DM could annihilate into two SM fermions if $m_V>m_\chi$ and
into $2V$ if $m_V<m_\chi$.
For the former case, $m_V$ is simply a free parameter used to control the strength of $\langle \sigma v\rangle$. The relation between $m_\chi$ and $m_V$ is obscured. However, for the later case,
we can approximate 
\begin{equation}
    \langle \sigma v\rangle = \frac{g_\chi^4}{16\pi m_\chi^2}\sqrt{1-\frac{m_V^2}{m_\chi^2}}\label{eq:DM_ann_crox}
\end{equation}
in terms of nonrelativistic (NR) DM. Thus, 
\begin{equation}
    \frac{m_V}{m_\chi} =\sqrt{1-\left(\frac{16\pi\langle \sigma v\rangle }{g_\chi^4}m_\chi^2\right)^2}.\label{eq:mv/mx}
\end{equation}
To produce the correct relic density, we carry $\langle \sigma v\rangle =6\times 10^{-26}~{\rm cm^3\,s^{-1}}$ and find
\begin{equation}
    \frac{m_V}{m_\chi} \sim 1 \label{eq:thermal_relic}
\end{equation}
holds in the interested range of $g_\chi$ and $m_\chi$ in this paper.\footnote{In Ref.~\cite{Lin:2021hen}, contributions from $t$ and $u$ channels to $\chi\bar{\chi}\to 2V$ are considered. A correction factor would be attached to Eq.~\eqref{eq:DM_ann_crox}. This does not change the conclusion, Eq.~\eqref{eq:thermal_relic}, but an analytical expression like Eq.~\eqref{eq:mv/mx} for $m_V/m_\chi$ is non-existent.}

On the other hand, the SIDM constraints \cite{Randall:2008ppe,Walker:2011zu,Boylan-Kolchin:2011lmk,Boylan-Kolchin:2011qkt,Elbert:2014bma,Tulin:2017ara,Adhikari:2022sbh} restrict $\sigma_{\chi\chi}$ in a band
\begin{equation}
    0.1~{\rm cm^2\,g^{-1}} \lesssim \sigma_{\chi\chi}/m_\chi \lesssim 10~{\rm cm^2\,g^{-1}}.\label{eq:SIDM_constraint}
\end{equation}
In the NR limit, we have
\begin{equation}
    \sigma_{\chi\chi}= \frac{g_\chi^4}{16\pi^2}\frac{m_\chi^2}{m_V^4}.\label{eq:sigxx}
\end{equation}
The lower bound of Eq.~\eqref{eq:SIDM_constraint} is much lose as it implies the 
DM self-interaction is too weak to be distinguished from collisionless DM.
Combining Eqs.~\eqref{eq:thermal_relic}-\eqref{eq:sigxx} and let $\eta\equiv 10~{\rm cm^2\,g^{-1}}$, we arrive
\begin{equation}
    m_V > \left(\frac{g_\chi}{2\sqrt{\pi}}\right)^{4/3}\eta^{-1/3}.
\end{equation}
Thus, the SIDM constraint sets a lower limit for $m_V$.

\section{BDM event numbers}\label{appx:BDM_numbers}
Here we show the BDM event number $N_s$ generated from $U(1)_{L_{\mu}-L_{\tau}}$ model for different $m_\chi$ with $\beta=0$ and $\pi$ in Super-K ($M_T=22.2$ kton) in Tab.~\ref{tab:Ns}. We choose $t_{\rm cut}=35\,$yrs,
$(T_{\chi,{\rm min}},T_{\chi,{\rm min}})=(5,100)\,$MeV
and $(g_V,g_\chi)=(10^{-5},10^{-2})$.
We take $R_s=8.5$~kpc and two different values of $\beta=0$ and $\pi$.
The values of the kinetic mixing $\varepsilon$ are labeled on top of the table.
For zero-kinetic mixing, the DM-$e$ interaction is induced by the
naturally arose parameter $\varepsilon^\prime = -g_V/70$.

\section{Identifying $t_p$ for model-agnostic case}\label{appx:tp_model_agnostic}

\begin{figure*}
\begin{centering}
\includegraphics[width=0.8\textwidth]{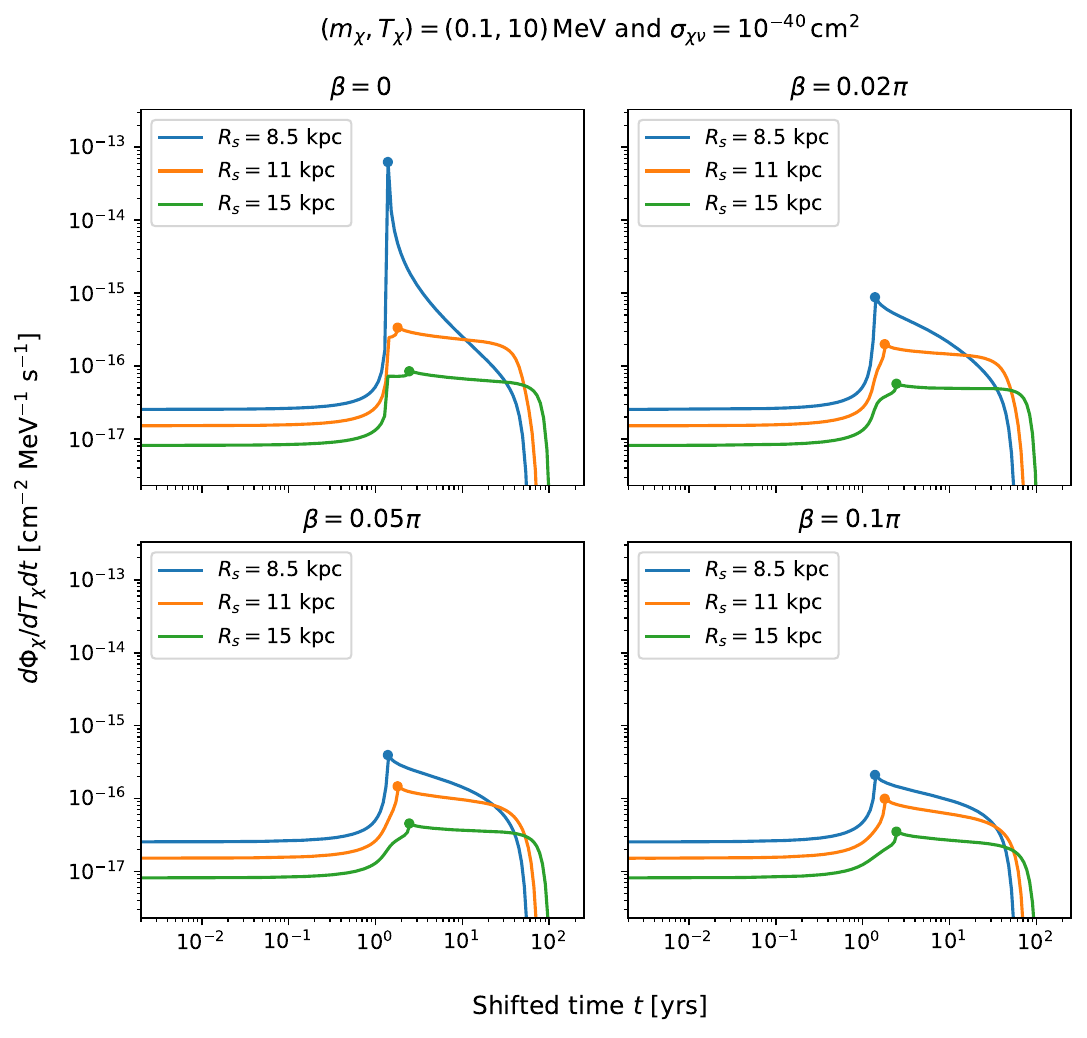}
\end{centering}
\caption{The BDM flux for various $\beta$ and $R_s$ in terms of model-agnostic. The dots indicate the corresponding $t_p$.
}
\label{fig:flux_Rs_v_beta}
\end{figure*}

In Fig.~\ref{fig:flux_Rs_v_beta} we display different BDM fluxes with $\beta=0,0.02\pi,0.04\pi$ and $0.5\pi$ with $R_s=8.5\,{\rm kpc}, 11\,{\rm kpc}$ and $15\,{\rm kpc}$ for the model-agnostic case. The DM-$\nu$ cross section is taken to be $\sigma_{\chi\nu}=10^{-40}\,{\rm cm}^2$.

For $\beta=0$ and $R_s=8.5$~kpc (SN located at GC) shown in the upper left panel, one sees that $t_p$ and the step wise increase of the BDM flux coincides. 
However, for the same $\beta=0$, the increase of the BDM fluxes for $R_s>8.5$~kpc occurs twice -- a sharper first one around $t=(8.5~{\rm kpc})(1/v_\chi-1/c)$ and a milder second one at $t\simeq t_p$.
Clearly, the first sharply-rising feature is associated with the increase of $n_\chi$ close to the GC, while the second rise around $t_p$ is related to the increase of the SN$\nu$ density $dn_\nu/dE_\nu$.
Thus, with larger $R_s$ and $\beta=0$, the sharply rising feature of the flux can occur earlier than $t_p$ for the model-agnostic scenario.
Consequently, it may be difficult to use $t_p$ to infer $m_\chi$ for these special SN locations.

This issue, however, is alleviated when considering non-zero but small $\beta$ values as shown in the other three panels. 
For these cases, since the line connecting the Earth and the SN location does not pass through GC, the BDM flux that arrived before $t_p$ with small scattering angle are less enhanced by the large $n_\chi$ around GC.
As a result, the sharply rising feature associated with the halo profile at GC diminishes. 
Instead, the main contribution to the smooth rising on the flux and turns at $t_p$ is due to $D\to0$ in $L_\nu$ in Eq.~\eqref{eq:dnv/dEv}.

For the $U(1)_{L_\mu-L_\tau}$ model, we verify that for the same values of $R_s$ and $\beta$ shown in Fig.~\ref{fig:flux_Rs_v_beta}, the sharp increase of the BDM flux is always associated with $t_p$, similar to what was discussed in Sec.~\ref{sec:res_SN_loc}. 
The difference in two different scenarios are related to the scattering angle dependence of the differential cross section $d\sigma_{\chi\nu}/d\cos\psi$ and can be understood as follows.  

For $\beta\ll 1$, the BDM flux at $t<t_p$ is dominated by those within a open angle $\theta\ll 1^\circ$, which obviously corresponds to small scattering angle $\psi\ll 1^\circ$ (see the blue triangle in Fig.~\ref{fig:geometry_anatomy}).
Numerically we found that  $d\sigma_{\chi\nu}/d\cos\psi\propto r^2$ when $\psi\ll 1^\circ$ where $r$ is the distance from the boosting location to GC (see also Fig.~\ref{fig:geometry_anatomy}).\footnote{This proportionality holds due to the constraint $t^\prime = D/c+d/v_\chi$ imposed in Eq.~\eqref{eq:BDM_flux}. This constraint conveys the $r$-dependence to $d\sigma_{\chi\nu}/d\cos\psi$ when determining the scattering angle $\psi$ at any boosted point.}
Thus, the $r^2$ dependence cancels the $r^{-2}$ divergence in the NFW profile when $r\to0$ and does not result in a sharp increase of BDM flux due to the increase of $n_\chi$ at GC.
On the other hand, the model-agnostic scenario has $d\sigma_{\chi\nu}/d\cos\psi=\sigma_{\chi\nu}\times f_\chi(\psi)$ approaching a constant when $\psi\to 0$ without an $r$ dependence. 
Consequently, there is no cancellation of the $r^{-2}$ divergence from NFW profile as $r\to 0$ that results in the first steep increase of BDM flux discussed above for $\beta=0$ and $R_s>8.5$ kpc.

Based on these results here, we conclude that only when a SN occurs at a location with $\beta\lesssim 0.02\pi$ and $R_s\sim 11$~kpc, the sharply rising feature of the BDM flux can depend on the underlying particle physics model and may not be associated with $t_p$. 
For other locations, the identification of $t_p$ based on the feature discussed in the main text and the use of it to along with $t_{\rm van}$, to infer $m_\chi$ in a model-independent way should still be possible.

\section{Sensitivity result for $T_{\chi,{\rm min}}=25$ MeV}\label{appx:Txmin_25MeV}

\begin{figure*}
\begin{centering}
\includegraphics[width=0.8\textwidth]{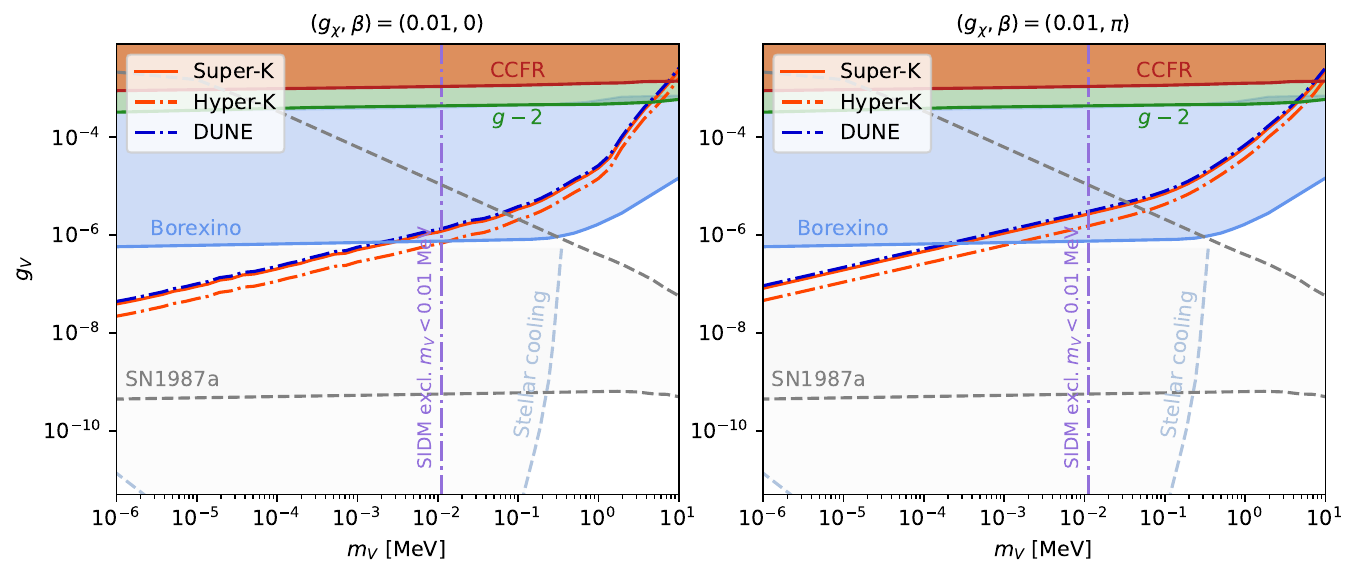}
\includegraphics[width=0.8\textwidth]{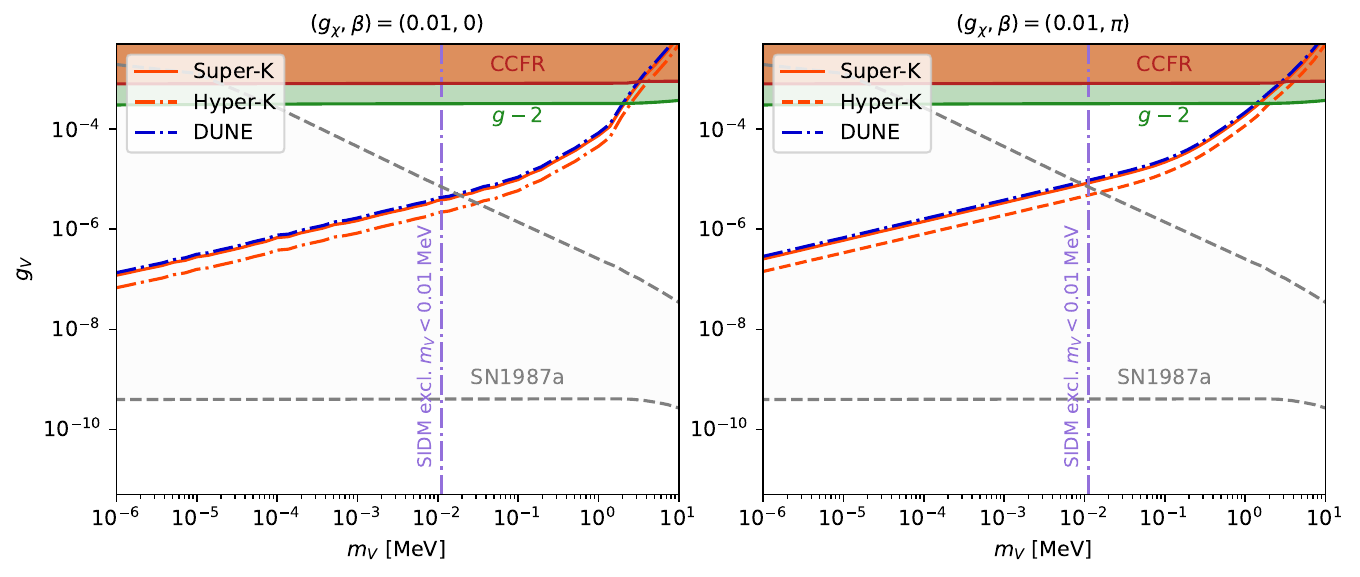}
\end{centering}
\caption{The notations follow those of  Figs.~\ref{fig:sensitivity_gx_equal_eps}  (upper row) and \ref{fig:sensitivity_gx_noeps} (lower row): The $U(1)_{L_\mu-L_\tau}$ charge $g_\chi=0.01$ for both rows and the threshold energy is lifted to $T_{\chi,{\rm min}}=25$ MeV.
}
\label{fig:sensitivity_Tx_min25MeV}
\end{figure*}

When taking $T_{\chi,{\rm min}}$ to be $25\,{\rm MeV}$ while retaining $T_{\chi,{\rm max}}= 100$~MeV, it removes the solar $\nu$ background. 
The only known background in this energy range is primarily the atmospheric $\nu$.
Thus, we can estimate $\mathcal{B}\approx 0.0059$ for water Cherenkov detector
and 0.0047 for liquid argon detector.
In principle, this may help identify more clean signatures from the SN$\nu$~BDM.
For instance, to a good approximation, $N_b$ is negligible and results in $N_s\approx 4$ in most of the parameter space we are interested.
However, as one can see from Fig.~\ref{fig:flux_v_Tx}, to get the total BDM events, we also have
to integrate over $T_\chi$ of interest.
If we set $T_{\chi,{\rm min}}=25$ MeV, then a significant part of lower energy BDM flux will not contribute to the event number. 
Thus, although the background is largely reduced, so do the signals.

We show the results for $g_\chi = 0.01$ in Fig.~\ref{fig:sensitivity_Tx_min25MeV}.
Except for  $T_{\chi,{\rm min}}=25$ MeV, other setups are identical to Figs.~\ref{fig:sensitivity_gx_equal_eps} and \ref{fig:sensitivity_gx_noeps}.
Compare this result to that obtained with $T_{\chi,{\rm min}}=5$ MeV shown in Fig.~\ref{fig:sensitivity_Tx_min25MeV}, one sees that the sensitivities are in fact weakened by a factor of few, due to
the loss of the BDM events from $T_\chi<25$ MeV.

\section{Code availability}
We provide a Python package \texttt{snorer} \cite{snorer2024}, which can fully reproduce the results in this paper. The package is available on both GitHub and PyPI. It offers numerous new features, such as including DM spikes, user-specified SN locations in arbitrary distant galaxies, and an implementation of any particle physics model. See its official page for further details.

%


\end{document}